\documentstyle[12pt]{article}
\def\vec#1{\ifcat#1\alpha\mbox{\boldmath $#1$}\else\mbox{\bf #1}\fi}
\newcommand{\bfrho}{\vec \rho}
\newcommand{\ltwid}{\raise.3ex\hbox{$<$\kern-.75em\lower1ex\hbox{$\sim$}}}
\newcommand{\gtwid}{\raise.3ex\hbox{$>$\kern-.75em\lower1ex\hbox{$\sim$}}}

\newcommand{\xq}{\begin{equation}}
\newcommand{\zq}{\end{equation}}
\newcommand{\beq}{\begin{eqnarray}}
\newcommand{\eeq}{\end{eqnarray}}

\title{{\hfill\normalsize NSF-ITP-94-74}\\[1.0cm]
Dynamical Flavor Symmetry Breaking by a Magnetic Field in $2+1$ Dimensions}

\author{
{\sl V.P. Gusynin$^1$}\\
{\sl V.A. Miransky$^{1,2}$}\\
{\sl I.A. Shovkovy$^1$}\\
{\sl $^1$Bogolyubov Institute for Theoretical Physics}\\
{\sl 252143 Kiev, Ukraine}\\
{\sl $^2$Institute for Theoretical Physics}\\
{\sl University of California, Santa Barbara, CA 93106-4030}
}
\date{}

\begin{document}

\maketitle

\begin{abstract}
It is shown that in $2+1$ dimensions, a constant magnetic field is a strong
catalyst of dynamical flavor symmetry breaking, leading to generating a fermion
dynamical mass even at the weakest attractive interaction between fermions. The
essence of this effect is that in a magnetic field, in $2+1$ dimensions, the
dynamics of fermion pairing is essentially one-dimensional. The effect is
illustrated in the Nambu-Jona-Lasinio model in a magnetic field. The low-energy
effective action in this model is derived and the thermodynamic properties of
the model are considered. The relevance of this effect for planar condensed
matter systems and for $3+1$ dimensional theories at high temperature
is pointed out.
\end{abstract}

\vfill

\section {Introduction}

Recently there has been considerable interest in relativistic field models in
$2+1$ dimensions. Besides that the sophisticated dynamics of these models is
interesting in itself, the models
also serve as effective theories for the description
of long wavelength excitations in planar condensed matter systems \cite{1,2}.
Also, their dynamics imitates the dynamics of $3+1$ dimensional theories at
high temperature.

In this paper, we will show that a constant magnetic field acts as a
strong catalyst of dynamical flavor symmetry breaking (generating fermion
masses) in $2+1$ dimensions. We will in particular show that there is a
striking similarity between the role of the magnetic field in
$(2+1)$-dimensional models and the role of the Fermi surface in the
Bardeen-Cooper-Schrieffer (BCS) theory of superconductivity \cite{3}: both of
them enhance the interactions of fermions in the infrared region (at small
momenta) thus leading to generating a fermion dynamical mass (energy gap in the
fermion spectrum) even at the weakest attractive interaction between fermions.

We note that necessity of a supercritical dynamics (with an effective coupling
constant $g$ being larger than
a critical value $g_c>0$) for generating fermion
dynamical masses is a common feature of the dynamics
 in $3+1$ and $2+1$ dimensions \cite{4}.
 As will be shown in this paper, in $2+1$ dimensions, a magnetic field reduces
the value of the critical coupling to zero.
We  note that the fact that a constant magnetic field enhances fermion
dynamical masses in the Nambu-Jona-Lasinio (NJL) model \cite{5} has been
already pointed out in Ref.~\cite{6}.
However, what we will show is not just that the magnetic field enhances the
dynamical mass created by the strong (supercritical) NJL interaction but that
in
$2+1$ dimensions, it catalyses generating the mass even at the weakest
attractive interaction. The essence of this effect is that in a magnetic field,
in $2+1$ dimensions, the dynamics of fermion pairing (relating essentially to
fermions at the lowest Landau level) is one-dimensional (see Sec.~2).

We stress that  this effect is universal, {\em i.e.,} model independent,
in $2+1$ dimensions.
This point may be important in connection with consideration of this effect
in such condensed matter phenomena as the quantum Hall effect \cite{1} and high
temperature superconductivity \cite{2}. Another, potentially
interesting,
application for this effect may be in $3+1$ dimensional theories at high
temperature (quark-gluon plasma in a magnetic field, for example): indeed, at
high temperature, their dynamics effectively reduces to the dynamics of $2+1$
dimensional theories.

As a soluble example, we shall consider the NJL model in a magnetic field, in
the leading order in $1/N$ expansion. We shall derive the low-energy effective
action in the model and also  study its thermodynamic properties.

The paper is organized as follows. In Secs.~2 and 3 we consider the problem of
a
relativistic fermion in a magnetic field in $2+1$ dimensions. We show that the
roots of the fact that a magnetic field is a strong catalyst of dynamical
flavor symmetry breaking in $2+1$ dimensions are actually in this problem. In
Secs.~4--7 we study the NJL model in a magnetic field in $2+1$
dimensions. We derive the low energy effective action and determine the
spectrum of long wavelength collective excitations in this model. In
Sec.~8 we
study the thermodynamic properties of the NJL model in a magnetic field. We
show that there is a symmetry restoring phase transition at high temperature.
In Sec.~9 we summarize the main results of the paper. In Appendices A and B,
some useful formulas and relations are derived. In Appendix C the reliability
of the $1/N$ expansion in this model is discussed.

\section { Dynamical Flavor Symmetry Breaking in the Problem of Fermions in
a Constant Magnetic Field}

In this section we will discuss the problem of relativistic fermions in a
magnetic field in $2+1$ dimensions. We will show that the roots of the fact
that a magnetic field is a strong catalyst of flavor symmetry
breaking are actually in this dynamics, which plays here the role similar to
that of the dynamics of the ideal Bose gas for an almost ideal Bose gas
in the theory of superfluidity \cite{4}.

The Lagrangian density in the problem of a relativistic fermion in a constant
magnetic field $B$ takes the following form in $2+1$ dimensions:
\xq
{\cal L} = \frac{1}{2} \left[\bar{\Psi}, (i\tilde{\gamma}^\mu
D_\mu-m)\Psi\right], \mu=0,1,2,    \label{eq:fre}
\zq
where the covariant derivative is
\xq
D_\mu=\partial_\mu-ie A^{ext}_\mu,\quad A^{ext}_\mu=-Bx_2\delta_{\mu1}.
\label{eq:dmu}
\zq
In $2+1$ dimensions, there are two inequivalent representations of the Dirac
algebra:
\xq
\tilde{\gamma}^0=\sigma_3, \tilde{\gamma}^1=i\sigma_1,
\tilde{\gamma}^2=i\sigma_2             \label{eq:pauli}
\zq
and
\xq
\tilde{\gamma}^0=-\sigma_3, \tilde{\gamma}^1=-i\sigma_1,
\tilde{\gamma}^2=-i\sigma_2             \label{eq:pauli1}
\zq
where $\sigma_i$ are Pauli matrices.

Let us begin by
 considering the representation (\ref{eq:pauli}).
 The energy spectrum in the problem (\ref{eq:fre}) depends on the sign of
 $eB$; let us first assume that $eB>0$. Then, the energy spectrum takes the
 form (to be concrete, we assume that $m\geq 0$) \cite{7}:
\beq
E_0 &=& \omega_0=m, \nonumber\\
E_n &=& \pm \omega_n=\pm\sqrt{m^2+2|eB|n},\ \  n=1,2,\dots .\label{eq:5}
\eeq
(the Landau levels).

The general solution is
\xq
\Psi(x)=\sum_{n,p} a_{np}u_{np}(x) + \sum_{n,p} b^+_{np} v_{n-p}(x),
\label{eq:sol}
\zq
where
\beq
u_{0p} &=& \frac{1}{(lL_1)^{1/2}} \exp (-i\omega_0t+ikx_1)
    \left[\begin{array}{c} w_0   (\xi) \\ 0     \end{array}\right],
 \nonumber\\
u_{np} &=& \frac{1}{(lL_1)^{1/2}} \exp (-i\omega_nt+ikx_1).\nonumber\\
&\cdot& \frac{1}{\sqrt{2\omega_n}}
        \left[ \begin{array}{c} \sqrt{\omega_n+m}  w_n(\xi) \\
               -i\sqrt{\omega_n-m}  w_{n-1}(\xi)
                 \end{array}\right], n\geq1,    \label{eq:7}
\\
v_{np} & = & \frac{1}{(lL_1)^{1/2}} \exp (i\omega_nt+ikx_1). \nonumber\\
&\cdot& \frac{1}{\sqrt{2\omega_n}} \left[\begin{array}{c} \sqrt{\omega_n-m}
        w_n(\xi) \\ i\sqrt{\omega_n+m}  w_{n-1}(\xi) \end{array}\right],
n\geq1.\nonumber
\eeq
Here $w_n(\xi)=(\pi^{1/2}2^nn!)^{-1/2}e^{-\xi^2/2}H_n(\xi)$, $H_n(\xi)$ are
Hermite polynomials,$l\equiv|eB|^{-1/2}$ is the magnetic length, $k=2\pi
p/L_1$ $(p=0,\pm1,\pm2,\dots)$, $L_1$ is the size in  the $x_1$-direction,
$\xi\equiv x_2/l+kl$. As $L_1\to\infty$, the density of the states at
each level $n$ is $|eB|/2\pi$ \cite{7}.

Thus the lowest Landau level with $n=0$ is special: while at $n\geq1$, there
are solutions corresponding to both fermion $(E_n=\omega_n)$ and antifermion
$(E_n=-\omega_n)$ states, the solution with $n=0$ describes only fermion
states.

As $eB\to-eB<0$, the solution becomes
\xq
\Psi(x) = \sum_{n,p} a_{np}v^c_{n-p}(x) + \sum_{np}b^+_{np} u^c_{np} (x)
\zq
where the charge conjugate spinors $v^c$ and $u^c$ are
$v^c=\tilde{\gamma}_2\bar{v}^T, \quad u^c=\tilde{\gamma}_2\bar{u}^T.$
Therefore at $eB<0$, the lowest Landau level with $n=0$ describes antifermion
states.

If we used the representation (\ref{eq:pauli1}) for Dirac's matrices, the
general
solution would be given by Eq.~(\ref{eq:sol}) with $u_{np}(x)$, $v_{np}(x)$
being
substituted by $(-1)^nv_{n-p}(-x),(-1)^nu_{n-p}(-x)$ (the factor $(-1)^n$
is introduced here for convenience):
\xq
\Psi(x)=\sum\limits_{n,p}c_{np}(-1)^nv_{n-p}(-x)+\sum\limits_{n,p}d_{np}
^\dagger(-1)^nu_{np}(-x) \label{eq:9}.
\zq
We note that the mass term in the Lagrangian density (\ref{eq:fre}) violates
parity
defined by
\xq
P: \Psi (x^0,x^1,x^2)\to\sigma_1\Psi(x^0,-x^1,x^2).
\zq
However if one uses the four-component fermions \cite{8}, connected with a
four-dimensional (reducible) representation of Dirac's matrices
\xq
\gamma^0= \left(\begin{array}{cc} \sigma_3 & 0 \\ 0&
-\sigma_3\end{array}\right)
, \gamma^1= \left(\begin{array}{cc}
i\sigma_1&0\\0&-i\sigma_1\end{array}\right),
\gamma^2= \left(\begin{array}{cc}i\sigma_2&0\\0&-i\sigma_2\end{array}\right),
\label{eq:11}
\zq
the mass term in the Lagrangian density
\xq
{\cal L} =\frac{1}{2} \left[\bar{\Psi}, \left(i\gamma^\mu
D_\mu-m\right)\Psi\right]\label{eq:free12}
\zq
preserves parity defined now as
\xq
P:\Psi (x^0,x^1,x^2)\to\frac{1}{i}\gamma^3\gamma^1\Psi (x^0,-x^1,x^2),
\zq
where the Dirac matrix $\gamma^3$ is
\xq
\gamma^3=i\left(\begin{array}{cc}0&I\\ I&0\end{array}\right).
\zq
The important point is that the Lagrangian density (\ref{eq:free12}) with $m=0$
is invariant
under the $U(2)$ (flavor) transformations with the generators
\xq
T_0=I,T_1=\gamma_5,T_2=\frac{1}{i}\gamma^3, T_3=\gamma^3\gamma^5,
\zq
where
\xq
\gamma^5=i\gamma^0\gamma^1\gamma^2\gamma^3=i
\left(\begin{array}{cc}0&I\\ -I&0\end{array}\right).
\zq
The mass term breaks this symmetry down to the $U(1)\times U(1)$ with  the
generators $T_0$ and $T_3$.

We note that the four-component fermions appear in low-energy effective actions
describing planar condensed matter systems with two sublattices \cite{2}.
Actually, usually they appear in the actions without the mass term, and the
important problem is to establish a criterion of dynamical flavor symmetry
breaking which may occur as a result of interaction between fermions
[4,9--16]. As was already indicated in Section~1, dynamical flavor
symmetry breaking in $2+1$ dimensions usually takes place only at a rather
strong effective coupling between fermions.

Let us now show that at $m=0$ and $B\neq 0$, the dynamical breakdown of the
$U(2)$ flavor symmetry takes place already in the theory (\ref{eq:free12}),
even without any
additional interaction between fermions. In order to prove this, we will show
that in the limit $m\to0$, the flavor condensate
$\langle0|\bar{\Psi}\Psi|0\rangle$ is nonzero:
$\langle0|\bar{\Psi}\Psi|0\rangle=-|eB|/2\pi$.

The condensate $\langle0|\bar{\Psi}\Psi|0\rangle$
 is expressed through the fermion propagator
$S(x,y)=$ \hfill\break
$=\langle0| T\Psi(x) \bar{\Psi} (y)| 0\rangle:$
\xq
\langle0|\bar{\Psi}\Psi|0\rangle=-\lim_{x\to y} trS(x,y).\label{eq:17}
\zq
The propagator $S$ is calculated (following the Schwinger (proper time)
approach \cite{16}) in Appendix~A. It is
\beq
S(x,y) &=&\exp (ie\int^x_y A_\lambda^{ext} dz^\lambda) \tilde{S} (x-y),
\label{eq:green}\\
\tilde{S} (x) &=& \int^\infty_0 \frac{ds}{8(\pi s)^{3/2}}
e^{-i(\frac{\pi}{4}+sm^2)} e^{-\frac{i}{4s}(x_\nu C^{\nu\mu}x_\mu)}.\nonumber\\
&\cdot& \Bigg[\left(m+\frac{1}{2s} \gamma^\mu C_{\mu\nu} x^\nu -
\frac{e}{2} \gamma^\mu F_{\mu\nu}^{ext} x^\nu\right) \Bigg( es B \cot (eBs) -
\nonumber \\
& & \qquad \quad -\frac{es}{2} \gamma^\mu\gamma^\nu F_{\mu\nu}^{ext}
\Bigg)\Bigg],
\eeq
where $C^{\mu\nu}=g^{\mu\nu}+ ((F^{ext})^2)^{\mu\nu}(1-eBs\cot(eBs))/B^2 $,
$F_{\mu\nu}^{ext}=\partial_\mu A^{ext}_\nu-\partial_\nu A^{ext}_\mu$
with $A_\mu^{ext}$ given in Eq.~(\ref{eq:dmu}). The integral in
Eq.~(\ref{eq:green}) is calculated
along the straight line.

The Fourier transform $\tilde{S}(k) = \int d^3x e^{ikx}\tilde{S} (x)$ is:
\beq
\tilde{S}(k) &=& \int^\infty_0 ds \exp\left[-ism^2+isk_0^2-is
\mbox{\boldmath $k$}^2
\frac{\tan(eBs)}{eBs} \right]\cdot \nonumber\\
&\cdot& \left[\left( \hat{k}+m+(k^2\gamma^1-k^1\gamma^2)\tan(eBs)\right)
\left(1+\gamma^1\gamma^2\tan(eBs)\right)\right] .\label{eq:20}
\eeq
Transferring this expression into Euclidean space $(k^0\to ik_3, s\to -is)$, we
find:
\beq
\tilde{S}_E(k) &=&-i \int^\infty_0 ds \exp \left[-s\Bigg(m^2+k^2_3+
\mbox{\boldmath$k$}^2
\frac{\tanh(eBs)}{eBs}\Bigg)\right]\label{eq:21} \\
&\cdot& \left[\left(-k_\mu\gamma_\mu+m+\frac{1}{i}(k_2\gamma_1-k_1\gamma_2)
\tanh(eBs)\right) \left(1+\frac{1}{i}\gamma_1\gamma_2\tanh(eBs)\right)\right]
\nonumber
\eeq
($\gamma_3=-i\gamma^0, \gamma_1\equiv\gamma^1, \gamma_2\equiv\gamma^2$ are
antihermitian matrices).

{}From Eqs.~(\ref{eq:17}), (\ref{eq:green}) and (\ref{eq:21}) we find the
following expression for the
condensate:
\beq
\langle0|\bar{\Psi}\Psi|0\rangle &=&-\frac{i}{(2\pi)^3} tr\int d^3k
\tilde{S}_E(k) = \nonumber\\
&=&- \lim_{\Lambda\to\infty}\lim_{m\to 0} \frac{4m}{(2\pi)^3} \int d^3k
\int^\infty_{1/\Lambda^2} ds \exp \left[-s\Bigg(m^2+k^2_3+
\mbox{\boldmath$k$}^2
\frac{\tanh(eBs)}{eBs}\Bigg)\right] = \nonumber\\
&=&-\lim_{\Lambda\to\infty}\lim_{m\to 0} \frac{m}{2\pi^{3/2}}
\int^\infty_{1/\Lambda^2} ds e^{-sm^2} (s^{-1/2}) (eB)\coth(eBs)=
\nonumber\\
&=&-\lim_{\Lambda\to\infty}\lim_{m\to 0} \frac{m}{2\pi^{3/2}}
\left(\pi^{1/2}|eB|\frac{1}{m}+O(\frac{1}{\Lambda}) \right) =-
\frac{|eB|}{2\pi} ,\label{eq:condensate}
\eeq
where $\Lambda$ is an ultraviolet cutoff.

Thus in a constant magnetic field, spontaneous breakdown of the flavor $U(2)$
symmetry takes place even though fermions do not acquire mass $(m=0)$! Note
that in $3+1$ dimensions, the result would be
$\langle0|\bar{\Psi}\Psi|0\rangle\sim mln m\to 0$ as $m\to0$.
Therefore
this is a specific $2+1$ dimensional phenomenon.

What is the physical basis of this phenomenon? In order to answer this
question, we note that a singular,$1/m$, behavior of the integral in
Eq.~(\ref{eq:condensate}) is formed at large, $s\to\infty$, distances ($s$ is
the proper time
coordinate). Actually one can see from Eq.~(\ref{eq:condensate}) that the
magnetic field
effectively removes the two space dimensions in the infrared region thus
reducing the dynamics to a one-dimensional dynamics which has much more
severe infrared singularities. From this viewpoint, the action of the
magnetic field in the present problem is similar to that of the Fermi surface
in the BCS theory \cite{3}.

This point is intimately connected with the form of the energy spectrum of
fermions in a constant magnetic field. Eqs.~(\ref{eq:5}), (\ref{eq:11})
and (\ref{eq:free12}) imply that for
the four-component fermions, the energy spectrum is:
\beq
E_0 &=& \pm \omega_0=\pm m, \nonumber\\
E_n &=& \pm \omega_n =\pm \sqrt{m^2+2|eB|n}, n\geq1.
\eeq
The density of the states with the energy $\pm\omega_0=\pm m$ is
$|eB|/2\pi$, and it is $|eB|/\pi$ at $n\geq1$. As $m\to0$, the
energy $E_0$
goes to zero and therefore there is the infinite vacuum degeneracy in this
case. The value of the condensate (\ref{eq:condensate}) is equal to the density
of the states at
the lowest Landau level.
This implies that spontaneous flavor symmetry breaking is intimately connected
with the dynamics of fermions at this level. In particular, since this
dynamics, described by one continuous variable $k_3=-ik^0$, is one-dimensional,
we get a simple explanation of the one-dimensional character of the dynamics of
flavor symmetry breaking in this problem. More precisely, the situation is the
following. Using the identity $tanh(x)=1-2exp(-2x)/[1+exp(-2x)]$
and the relation \cite{22}:
\begin{equation}
(1-z)^{-(\alpha+1)}\exp\left(\frac{xz}{z-1}\right)=\sum_{n=0}^{\infty}
L_n^{\alpha}(x)z^n,
\end{equation}
where $L_n^{\alpha}(x)$ are the generalized Laguerre polynomials, the
propagator $\tilde{S}_E (k)$ can be decomposed over the Landau level
poles \cite{Cho}:
\begin{equation}
\tilde{S}_E (k)=-iexp\left(-\frac{{\bf k}^2_{\perp}}{|eB|}\right)
\sum_{n=0}^{\infty}(-1)^n\frac{D_n(eB,k)}{k_3^2+m^2+2|eB|n}\label{eq:poles}
\end{equation}
with
\begin{eqnarray}
D_n(eB,k)&=&(m-k_3\gamma_3)\bigg[(1-i\gamma_1\gamma_2sign(eB))
L_n(2\frac{{\bf k}^2_{\perp}}{|eB|})\nonumber\\
&-&(1+i\gamma_1\gamma_2sign(eB))L_{n-1}
(2\frac{{\bf k}^2_{\perp}}{|eB|})\bigg]
+4(k_1\gamma_1+k_2\gamma_2)L_{n-1}^1(2\frac{{\bf k}^2_{\perp}}{|eB|}),
\end{eqnarray}
where $L_n \equiv L_n^0$ and $L_{-1}^{\alpha}(x)=0$ by definition. Then
Eq.(\ref{eq:poles}) implies that as $m\to 0$, the condensate appears
due to the lowest Landau level:
\beq
\langle0|\bar{\Psi}\Psi|0\rangle\simeq-\frac{m}{2\pi^3} \int d^3k
\frac{\exp\left(-{\bf k}^2_{\perp}/|eB|\right)}{k_3^2+m^2}=
-\frac{|eB|}{2\pi}.
\eeq
We would like to note that in $3+1$ dimensions, the dynamics at the
lowest Landau level, described by two continuous variables $k_3$ and $k_4$,
is two dimensional\footnote{The fact that a magnetic field reduces the
effective dimension of the dynamics of the fermion pairing by two units
was observed ealier in the theory of superconductivity \cite{17}.}.
Actually, the lowest Landau level pole in $\tilde{S}_E(k)$ is in $3+1$
dimensions:
\beq
\tilde{S}^{0}(k)=-i\exp\left(-\frac{{\bf k}^2_{\perp}}{|eB|}\right)
\frac{m-k_4\gamma_4-k_3\gamma_3}{k_4^2+k_3^2+m^2}(1-i\gamma_1\gamma_2).
\eeq
The contribution of this level to the condensate is now:
\beq
\langle0|\bar{\Psi}\Psi|0\rangle\simeq -|eB|\frac{m}{4\pi^2}
\left(\ln\frac{\Lambda^2}{m^2}+O(m^0)\right)\stackrel{m\to 0}{\rightarrow} 0,
\eeq
{\em i.e.}, there is no spontaneous flavor symmetry breaking in a magnetic
field in $3+1$ dimentions.

In the next section, we will discuss aspects of spontaneous flavor symmetry
breaking for ($2+1$)--dimentional fermions in a magnetic field in more detail.

\section{More About Dynamical Flavor Symmetry Breaking for Fermions
in a Magnetic Field}

As was shown in the preceding section, the flavor condensate
$\langle 0|\bar{\Psi}\Psi|0\rangle$ is nonzero as the fermion mass $m$
goes to zero. Although usually this fact is considered as a firm signature
of spontaneous flavor (or chiral) symmetry breaking, the following questions
may arise in this case:

{\em a)} Unlike the conventional spontaneous flavor (chiral) symmetry
breaking, the dynamical mass of fermions equals zero in this problem.
Is spontaneous flavor symmetry breaking "real" in this case?

{\em b)} The vacuum $|0\rangle$ was defined as $\lim_{m\to 0}|0\rangle_m$ of
the vacuum $|0\rangle_m$ in the theory with $m\neq 0$. The vacuum $|0\rangle$
corresponds to a particular filling of the lowest Landau level. Indeed, at
$m\neq 0$, in the vacuum $|0\rangle_m$, the states with $E_0=m>0$ are empty
and the states with $E_0=-m$ are filled, {\em i.e.} the vacuum
$|0\rangle=\lim_{m\to 0}|0\rangle_m$ is annihilated by all the operators
$a_{0p}$, $d_{0p}$ and $a_{np}$, $b_{np}$, $c_{np}$, $d_{np}$ ($n\geq 1$).
On the other hand, at $m=0$, there is the infinite degeneracy of the vacuum
in this problem, connected with different fillings of the lowest Landau level.
Why should one choose the filling leading to the vacuum $|0\rangle$? And is
there a filling of the lowest Landau level leading to the ground state which
is invariant under the flavor $U(2)$? One might think that the latter
possibility  would imply that spontaneous flavor symmetry breaking can be
avoided.

In this section we will show that there is a genuine realization of the
spontaneous breakdown of the flavor symmetry in the present problem. More
precisely, we shall show that this phenomenon satisfies all the criterions
of the spontaneous symmetry breaking phenomenon established by Haag long ago
\cite{Haag}. We will also discuss such related questions as the status of
Nambu--Goldstone (NG) modes and induced quantum numbers \cite{Niemi,Kovner}
in this problem.

Let us begin by constructing the charge operators
$Q_i=1/2\int d^2x[\Psi^{\dagger}(x),T_i\Psi(x)]$ of the flavor $U(2)$ group.
By using Eqs.(\ref{eq:sol}), (\ref{eq:7}) and (\ref{eq:9}), we find:
\beq
Q_0=\sum\limits_p\left(a_{0p}^\dagger a_{0p}-d_{0-p}^\dagger d_{0-p}\right)
+\sum\limits_{n=1}^\infty\sum\limits_p\left(a_{np}^\dagger a_{np}
-b_{np}^\dagger b_{np}+c_{np}^\dagger c_{np} -d_{np}^\dagger  d_{np}\right)
\nonumber
\eeq
\beq
Q_1=i\sum\limits_p\left(a_{0p}^\dagger d_{0-p}^\dagger -d_{0-p}a_{0p}\right)
+i\sum\limits_{n=1}^\infty\sum\limits_p\left(a_{np}^\dagger c_{np}-c_{np}^
\dagger a_{np}+b_{np}^\dagger d_{np}-d_{np}^\dagger b_{np}
\right),\nonumber
\eeq
\beq
Q_2=\sum\limits_p\left(a_{0p}^\dagger d_{0-p}^\dagger +d_{0-p}a_{0p}\right)
+\sum\limits_{n=1}^\infty\sum\limits_p\left(a_{np}^\dagger c_{np}+c_{np}^
\dagger a_{np}+b_{np}^\dagger d_{np}+d_{np}^\dagger b_{np}
\right),     \label{eq:charge}
\eeq
\beq
Q_3=\frac{|eB|}{2\pi}S+\sum\limits_p\left(a_{0p}^\dagger a_{0p}
+d_{0-p}^\dagger  d_{0-p}\right)+\sum\limits_{n=1}^\infty\sum\limits_p
\left(a_{np}^\dagger a_{np}-b_{np}^\dagger b_{np}-c_{np}^\dagger c_{np}+
d_{np}^\dagger d_{np}\right)
\nonumber
\eeq
where $a_{np},c_{np},(b_{np},d_{np})$ are annihilation operators of
fermions (antifermions) from the n-th Landau level and $S=L_1L_2$ is the
$2$--dimensional volume. Now we can construct a set of the degenerate
vacua:
\beq
|\theta_1,\theta_2\rangle=\exp(iQ_1\theta_1+iQ_2\theta_2)|0\rangle
\eeq
where, we recall, the vacuum $|0\rangle=\lim_{m\to 0}|0\rangle_m$
is annihilated by all the operators $a_{np},b_{np},c_{np}$  and $d_{np}$ .
As one can see from Eq.(\ref{eq:charge}), the crucial point for the existence
of the
continuum set of the degenerate vacua is the first sum, over the states
at the lowest Landau level, in the charges $Q_1$ and $Q_2$.

The presence of such a set of the degenerate vacua is a signal of the
spontaneous breakdown, $U(2)\to U(1)\times U(1)$. Note that the vacua
$|\theta_1,\theta_2\rangle$ can be also constructed by replacing the mass
term $m\bar{\Psi}\Psi$ by
$m\bar{\Psi}_{\theta_1,\theta_2}\Psi_{\theta_1,\theta_2}$, where
$\Psi_{\theta_1,\theta_2}=\exp(iQ_1\theta_1+iQ_2\theta_2)\Psi$, and then
performing the limit $m\to 0$. Again, this is a standard way of constructing
degenerate vacua in the case of spontaneous breakdown of a symmetry.

One can check that different vacua $|\theta_1,\theta_2\rangle$ become
orthogonal as size $L_1$ in the $x_1$--direction goes to infinity. For
example:
\beq
|\langle 0,\theta_2|0,\theta_2^\prime\rangle|=\prod\limits_p
|\cos\theta|=\exp(L_1\int dk\ln|\cos\theta|),\quad\theta=\theta_2^\prime-
\theta_2,\nonumber
\eeq
and at $\theta\neq0$ or $\pi$ , it goes to zero as $L_1\to\infty$
( the vacuum $|0,\theta_2+\pi\rangle=-|0,\theta_2\rangle$). It also goes to
zero as the maximum momentum $|k_{max}|=\Lambda$ ( $\Lambda$ is an ultraviolet
cutoff) goes to infinity. As usual, this point reflects the fact that
spontaneous
symmetry breaking occurs only in a system with an infinite number of degrees
of freedom. One can check that in this case all states (and not just vacua)
from different Fock spaces $F_{\{\theta_1\theta_2\}}$, defined by different
vacua $|\theta_1,\theta_2\rangle$,
are orthogonal. That is, different vacua $|\theta_1,\theta_2\rangle$ define
nonequivalent representations of canonical commutation relations.

On the other hand, taking the ground state
\beq
|\Omega\rangle=C\int d\mu(\theta_1,\theta_2,\theta_3)
|\theta_1,\theta_2\rangle,\label{eq:omega}
\eeq
where $d\mu$ is the invariant measure of $SU(2)$ and $C$ is a normalization
constant, we are led to the vacuum $|\Omega\rangle$ which is a singlet
with respect to the flavor $U(2)$. In fact, the set of the vacua
$\{|\theta_1,\theta_2\rangle\}$ can be decomposed in irreducible
representations of $SU(2)$:
\beq
\{|\theta_1,\theta_2\rangle\}=\{|\Omega^{(i)}\rangle\}
\eeq
Why should we consider the vacua $|\theta_1,\theta_2\rangle $ instead of the
vacua $|\Omega^{(i)}\rangle$?

To answer to this question, we consider, following Haag \cite{Haag}, the
clusterization property of Green's functions. It means the following. Let us
consider a Green's function
\beq
G^{(n+k)}=\langle 0|T\prod_{i=1}^{n}A_i(x_i)\prod_{j=1}^{k}B_j(y_j)|0\rangle,
\eeq
where $A_i(x_i)$, $B_j(y_j)$ are some local operators. The clusterization
property implies that when $r^2_{ij}\to \infty$ ($r^2_{ij}=
({\bf x}_i-{\bf y}_j)^2$) for all $i$ and $j$, the Green's
function then factorizes as follows:
\beq
G^{(n+k)}\to \langle 0|T\prod_{i=1}^{n}A_i(x_i)|0\rangle
\langle 0|T\prod_{j=1}^{k}B_j(y_j)|0\rangle.
\eeq
The physical meaning of this property is clear: clusterization implies the
absence of instantaneous long--range correlations in the system, so that the
dynamics in two distant spatially--separated regions are independent.

The clusterization property takes place for all the vacua
$|\theta_1,\theta_2\rangle $. The simplest way to show this is to note that
the vacuum $|\theta_1,\theta_2\rangle $ appears in the limit $m\to 0$ from
the vacuum in the system with the mass term
$m\bar{\Psi}_{\theta_1\theta_2}\Psi_{\theta_1\theta_2}$. Since at $m\neq 0$,
the vacuum in this system is unique, the clusterization is valid at
{\em every} value of $m\neq 0$. Therefore it is also valid in the limit
$m\to 0$, as far as the Green's functions exist in this limit.
In connection with that, we would like to note that, in thermodynamic limit
$L_1,L_2\to \infty$, the vacuum $|\theta_1,\theta_2\rangle $ is the only
normalizable and translation invariant state in the Fock space
$F_{\theta_1\theta_2}$. To show this, let us introduce the
operators $a_n(k)=(L_1/2\pi)^{1/2}a_{np}$, $b_n(k)=(L_1/2\pi)^{1/2}b_{np}$,
$c_n(k)=(L_1/2\pi)^{1/2}c_{np}$, $d_n(k)=(L_1/2\pi)^{1/2}d_{np}$, where
$k=2\pi p/L_1$. They satisfy the commutation relations $[a_{n}(k),
a_{n^\prime}^{\dagger}(k^{\prime})]=\delta_{nn^{\prime}}\delta(k-k^{\prime})$,
etc. Therefore, though states of the form
$\prod_{i}a^{\dagger}_0(k_i)\prod_{j}d^{\dagger}_0(k_j)
|\theta_1,\theta_2\rangle $ have zero energy, they are not normalizable and,
at $\sum_{i}k_i+\sum_{j}k_j\neq 0$, not translation invariant.

On the other hand, the clusterization property is valid not for all Green's
functions on the vacua $|\Omega^{(i)}\rangle$. As an example, consider the
Green's function
\beq
G^{(4)}= \langle \Omega|T(\bar{\Psi}(x_1)\Psi(x_2))
(\bar{\Psi}(y_1)\Psi(y_2))|\Omega\rangle,
\eeq
where $|\Omega\rangle$ is the vacuum singlet (\ref{eq:omega}).
Since the bilocal operator $\bar{\Psi}(x_1)\Psi(x_2)$ is assigned to the
triplet of $SU(2)$, the clusterization property would imply that
\beq
G^{(4)}\to \langle \Omega|T(\bar{\Psi}(x_1)\Psi(x_2))|\Omega\rangle
\langle\Omega|(\bar{\Psi}(y_1)\Psi(y_2))|\Omega\rangle\to 0.
\eeq
as $r^2_{ij}=({\bf x}_i-{\bf y}_j)^2\to \infty$. However,
since
\begin{eqnarray}
\langle\Omega|T(\bar{\Psi}(x_1)\Psi(x_2))|\Omega^{(3)}\rangle &\neq&
0,\nonumber\\
\langle\Omega^{(3)}|T(\bar{\Psi}(y_1)\Psi(y_2))|\Omega\rangle &\neq& 0,
\end{eqnarray}
where $|\Omega^{(3)}\rangle$ is a state from the vacuum triplet, we see
that $G^{(4)}$ does not vanish as $r_{ij}^2\to \infty$.

Thus the clusterization property does not take place for the
$|\Omega^{(i)}\rangle$--vacua.

This is a common feature of the systems with spontaneous continuous symmetry
breaking \cite{4,Haag}: an orthogonal set of vacua can either be labeled
by the continuous parameters $\{\theta_i\}$, connected with the generators
$Q_i$ of the broken symmetry, or it can be decomposed in irreducible
representations of the initial group. However, the latter vacua do not
satisfy the clusterization property.

All the Fock spaces $F_{\{\theta_1\theta_2\}}$ yield physically
equivalent descriptions
of the dynamics: in the space $F_{\{\theta_1\theta_2\}}$, the $SU(2)$
spontaneously breaks down to $U_{\{\theta_1\theta_2\}}(1)$, where the
$U_{\{\theta_1\theta_2\}}(1)$  symmetry is connected with the generator
$Q_3^{\{\theta_1\theta_2\}}=\exp(iQ_1\theta_1+iQ_2\theta_2)Q_3
\exp(-iQ_1\theta_1-iQ_2\theta_2)$. Are there NG modes in the present
system? To answer to this question, let us
consider the thermodynamic limit $L_1,L_2\to \infty$.
 One can see that in every Fock space $F_{\{\theta_1\theta_2\}}$, with
the vacuum $|\theta_1, \theta_2\rangle$, there are a lot of "excitations"
with nonzero momentum $k$ and zero energy $E$ created by the operators
$a_{0}^\dagger(k)$ and $d_{0}^\dagger(k)$. However there are no
genuine ( i.e. with a nontrivial dispersion law) NG modes:
the energy $E$ is $E\equiv0$ at the lowest Landau level. Since the Lorentz
symmetry is broken by a magnetic field, this point does not contradict to
Goldstone's theorem\footnote{As to a nonrelativistic analog of the
Goldstone theorem, it has been proved only for
translation invariant systems with finite range interactions
\cite{18}. Systems in a magnetic field do not satisfy this condition.}.
This of course does not imply that the existence of NG modes is incompatible
with a magnetic field: the situation is model dependent. As will be shown in
Secs.4-7, even the weakest attractive interaction in the problem of $2+1$
dimensional fermions in a magnetic field is enough to "resurrect" the genuine
NG modes. The key point for their existence is that the flavor condensate
$\langle 0 |\bar{\Psi} \Psi | 0\rangle$ and the NG modes are neutral, and the
translation symmetry in neutral channels is not violated by a magnetic
field (see the next section).
We shall also see that the "excitations" from the lowest Landau
level (with quantum numbers of the NG modes) in the problem of free fermions
in a magnetic field can be interpreted as "remnants" of the genuine NG modes
in the limit when the interaction between fermions is being switched off.
Moreover, we shall see in Sec.5 that the vacua $|\theta_1,\theta_2\rangle$
constructed above yield a very good approximation for the vacua
of  systems with weakly interacting fermions in a magnetic field (in fact,
it appears that the role of the vacua $|\theta_1,\theta_2\rangle$
is the same as that of the $\theta$--vacua of the ideal Bose gas for an
almost ideal Bose gas in the theory of superfluidity \cite{4}).

In conclusion, let us discuss the phenomenon of induced quantum numbers
\cite{Niemi,Kovner} in this problem.
As it follows from Eq.(\ref{eq:charge}), the vacuum
$|\theta_1, \theta_2\rangle$ is an eigenstate of the density operator
$\rho_3^{\{\theta_1 \theta_2\}}=\lim_{S\to \infty}
Q_3^{\{\theta_1 \theta_2\}}/S$ with a nonzero value:
\beq
\rho_3^{\{\theta_1 \theta_2\}}|\theta_1, \theta_2\rangle=
\frac{|eB|}{2\pi}|\theta_1, \theta_2\rangle.
\eeq
Thus, there is the induced quantum number of the operator
$\rho_3^{\{\theta_1 \theta_2\}}$ in the $|\theta_1 \theta_2\rangle$
vacuum\footnote{Note that this fact agrees with the consideration in
Ref.\cite{Kovner}.}.
This fact is intimately connected with the phenomenon of spontaneous
flavor symmetry breaking in this problem. Indeed, since
$Q_3^{\{\theta_1 \theta_2\}}$ is one of the
generators of the non--Abelian $SU(2)$ symmetry, its vacuum eigenvalue would
be equal to zero if the symmetry were exact and the vacuum were assigned to
the singlet (trivial) representation of $SU(2)$. This is in contrast to the
case of Abelian $U(1)$ symmetry: since $U(1)$ has an infinite
number of one dimensional representations, the vacuum can be an eigenstate
of the charge density $\rho=\lim_{S\to \infty}Q/S$ with an arbitrary
eigenvalue in that case.

Note that, since the $SU(2)$ is spontaneously broken here, it is appropriate
to redefine the generator of the exact $U_{\{\theta_1\theta_2\}}(1)$
symmetry as $\tilde{Q}_3^{\{\theta_1\theta_2\}}=Q_3^{\{\theta_1\theta_2\}}
-|eB|S/2\pi$.

\section {The Nambu-Jona-Lasinio Model in a Magnetic Field. General
Consideration.}

In this and the following four sections, we shall consider the NJL model
in $2+1$ dimensions. This model gives a clear illustration of the general
fact that a constant magnetic field is a strong catalyst of generating a
fermion dynamical mass in $2+1$ dimensions.

Let us consider the $2+1$ dimensional NJL model invariant under the
$U(2)$ flavor transformations:
\beq
{\cal L} &=& \frac{1}{2} \left[\bar{\Psi}, i\gamma^\mu D_\mu\Psi\right]
+ \frac{G}{2} \Bigg[(\bar{\Psi}\Psi)^2+(\bar{\Psi}i\gamma_5 \Psi)^2 +
\nonumber\\
&+& (\bar{\Psi}\gamma_3\Psi)^2\Bigg],   \label{eq:lag1}
\eeq
where $D_\mu$ is the covariant derivative (\ref{eq:dmu}) and fermion fields
carry an additional, ``color", index $\alpha=1,2,\dots, N$. This theory
is equivalent to a theory with the Lagrangian density
\beq
{\cal L} &=& \frac{1}{2} \left[\bar{\Psi}, i\gamma^\mu D_\mu\Psi\right] -
\bar{\Psi} \left(\sigma+\gamma^3\tau+i\gamma^5\pi\right)
\Psi - \nonumber\\
&-& \frac{1}{2G} \left(\sigma^2+\pi^2+\tau^2\right).\label{eq:lag2}
\eeq
The Euler-Lagrange equations for the auxiliary fields $\sigma,\tau$ and $\pi$
take the form of constraints:
\xq
\sigma=-G(\bar{\Psi}\Psi), \tau=-G(\bar{\Psi}\gamma^3\Psi), \pi=-G
(\bar{\Psi}i\gamma^5\Psi).  \label{eq:cons}
\zq
The Lagrangian density (\ref{eq:lag2}) reproduces Eq.~(\ref{eq:lag1})
upon application of the constraints (\ref{eq:cons}).

The effective action for the composite fields is expressed through the path
integral over fermions:
\beq
\Gamma(\sigma,\tau,\pi) &=&-\frac{1}{2G} \int d^3x(\sigma^2+\tau^2+\pi^2) +
   \tilde{\Gamma}(\sigma,\tau,\pi), \label{eq:effact}\\
\exp(i\tilde{\Gamma}) &=& \int [d\Psi][d\bar{\Psi}] \exp \Bigg\{
\frac{i}{2} \int
d^3x \Bigg[ \bar{\Psi}, \Bigg\{ i\gamma^\mu D_\mu - \nonumber\\
&-& (\sigma+\gamma^3\tau+i\gamma^5\pi) \Bigg\} \Psi\Bigg]\Bigg\}=
 \exp \Bigg\{ Tr Ln \Bigg[
i\gamma^\mu D_\mu - \nonumber\\
\qquad\qquad&-& (\sigma+\gamma^3\tau+i\gamma^5\pi)\Bigg]\Bigg\},\label{eq:exp}
\eeq
i.e.,
\xq
\tilde{\Gamma}(\sigma,\tau,\pi) =- i TrLn \left[i\gamma^\mu D_\mu-
(\sigma+\gamma^3\tau+i\gamma^5\pi)\right].\label{eq:tildefact}
\zq
As $N\to\infty$, the path integral over the composite (auxiliary) fields is
dominated by stationary points of the action:
$\delta\Gamma/\delta\sigma
=\delta\Gamma/\delta\tau
=\delta\Gamma/\delta\pi=0$. We will analyze the dynamics in this limit
by using the expansion of the action $\Gamma$ in powers of derivatives of the
composite fields.

Is the $1/N$ expansion reliable in this problem? This question appears
naturally since, as was emphasized in Sec.2, a magnetic field reduces the
dimension of the dynamics of the fermion pairing by two units.If such a
reduction took place for the whole dynamics (and not just for that of the
fermion pairing), the $1/N$ perturbative expansion would be unreliable. In
particular the contribution of the NG modes in the gap equation, in
next--to--leading order in $1/N$, would lead to infrared divergences.
Just such a
situation takes place in the $1+1$ dimensional Gross-Neveu model with a
continuous chiral symmetry ~\cite{19}. This phenomenon reflects the
Mermin-Wagner-Coleman theorem ~\cite{20} forbidding spontaneous breakdown
of continuous symmetries in space dimensions lower than two.

Fortunately, as will be shown in Appendix C, this  is not the case in the
present problem. The central point is that condensate
$\langle 0|\bar{\Psi}\Psi|0\rangle$ and the NG modes are neutral in this
problem. As we shall see in Sec.6 and Appendix C, this is
reflected in the structure of the propagator of the NG modes : unlike the
fermion propagator, it has a genuine $2+1$ dimensional structure. As a
result, their contribution to the dynamics does not lead to infrared
divergences, and the $1/N$ expansion is reliable in this problem. This
point is intimately connected with the status of the space-translation
symmetry in a constant magnetic field. In the gauge (\ref{eq:dmu}), the
translation
symmetry along the $x_2$-direction is broken (though it can be restored by
applying a certain gauge transformation). Therefore the momentum $k_2$ is
a bad quantum number for fermions and all other charged states ( see Eqs.
(\ref{eq:sol}) and (\ref{eq:7})). However, for neutral states , both the
momenta $k_1$ and $k_2$ of their center of mass are conserved quantum
numbers (this property is gauge-invariant) \cite{Zak}. In order to
show this fact in the gauge (\ref{eq:dmu}),
let us introduce the following operators describing space translations in
first quantized theory:
\beq
\hat{P}_{x_1}=\frac{1}{i}\frac{\partial}{\partial x_1}, \qquad
\hat{P}_{x_2}=\frac{1}{i}\frac{\partial}{\partial x_2}+\hat{Q}Bx_1,
\label{eq:tran}
\eeq
where $\hat{Q}$ is the charge operator. One can easily check that these
operators commute with the Hamiltonian of the Dirac equation in a constant
magnetic field. Also, the commutator $[\hat{P}_{x_1},\hat{P}_{x_2}]$ is:
\beq
[\hat{P}_{x_1},\hat{P}_{x_2}]=-i\hat{Q}B.
\eeq
Therefore the commutator equals zero for neutral states, and both the momenta
$k_1$ and $k_2$ can be used to describe the dynamics of the center of mass of
neutral states. As we shall see, this point is important for providing the
$(2+1)$--dimensional character of this dynamics.

\section{The NJL Model in a Magnetic Field. The Effective Potential.}

We begin the calculation of $\Gamma$ by calculating the effective potential
$V$. Since $V$ depends only on the $SU(2)$-invariant
$\rho^2=\sigma^2+\tau^2+\pi^2$, it is sufficient to consider a configuration
with $\tau=\pi=0$ and $\sigma$ independent of $x$. So now
$\tilde{\Gamma}(\sigma)$ is
\xq
\tilde{\Gamma}(\sigma) =- i TrLn (i\hat{D} -\sigma)=-iLnDet(i\hat{D}-\sigma),
\zq
where $\hat{D}\equiv\gamma^\mu D_\mu$. Since
\xq
Det(i\hat{D}-\sigma)=Det(\gamma^5(i\hat{D}-\sigma)\gamma^5)=Det(-i\hat{D}
  -\sigma),
\zq
we find that
\beq
\tilde{\Gamma}(\sigma) &=&-\frac{i}{2} Tr\left[Ln(i\hat{D}-\sigma)+Ln
(-i\hat{D}-\sigma)\right] = \nonumber\\
&=&- \frac{i}{2} TrLn (\hat{D}^2+\sigma^2).
\eeq
Therefore $\tilde{\Gamma}(\sigma)$ can be expressed through
the following  integral over  the
proper time $s$:
\xq
\tilde{\Gamma}(\sigma)=-\frac{i}{2}TrLn (\hat{D}^2+\sigma^2)=\frac{i}{2} \int
d^3x\int^\infty_0\frac{ds}{s}tr
\langle x|e^{-is(\hat{D}^2+\sigma^2)}|x\rangle  \label{eq:gamm}
\zq
where
\beq
\hat{D}^2 &=& D_\mu D^\mu - \frac{ie}{2} \gamma^\mu\gamma^\nu F_{\mu\nu}^{ext}
= \\
\qquad &=& D_\mu D^\mu + ie \gamma^1\gamma^2 B. \nonumber
\eeq
The matrix element $\langle x|e^{-is(\hat{D}^2+\sigma^2)}|y\rangle$ can be
calculated by using the Schwinger approach \cite{16}. It is:
\beq
\langle x|e^{-is(\hat{D}^2+\sigma^2)}|y\rangle &=& e^{-is\sigma^2}
\langle x| e^{-isD_\mu D^\mu}|y\rangle \Bigg[ \cos (eBs)+ \nonumber\\
\qquad&+& \gamma^1\gamma^2 \sin (eBs)\Bigg]= \label{eq:heatker}\\
&=& \frac{e^{-i\frac{\pi}{4}}}{8(\pi s)^{3/2}} e^{-i(s\sigma^2-S_{cl})}
  \Bigg[eBs \cot(eBs)+\gamma^1\gamma^2 eBs\Bigg] \nonumber
\eeq
where
\beq
S_{cl} &=& e\int^x_y A^{ext}_\lambda dz^\lambda - \frac{1}{4s} (x-y)_\nu
   \Bigg( g^{\nu\mu}+\frac{(F^2_{ext})^{\nu\mu}}{B^2} \cdot \nonumber\\
&\cdot& \left[1-eBs\cot(eBs)\right]\Bigg) (x-y)_\mu.
\eeq
Here the integral $\int^x_y A^{ext}_\lambda
 dz^\lambda$ is taken along the straight
line.

Substituting Eq.~(\ref{eq:heatker}) in Eq.~(\ref{eq:gamm}), we find
\xq
\tilde{\Gamma}(\sigma) = \frac{iNe^{-i\frac{\pi}{4}}}{4\pi^{3/2}} \int d^3x
\int^\infty_0 \frac{ds}{s^{5/2}} e^{-is\sigma^2} eBs \cot (eBs).
\zq
Therefore the effective potential is
\xq
V(\rho)= \frac{\rho^2}{2G}+\tilde{V}(\rho)= \frac{\rho^2}{2G}
+\frac{N}{4\pi^{3/2}} \int^\infty_{1/\Lambda^2} \frac{ds}{s^{5/2}}
e^{-s\rho^2}eBs \coth (eBs),             \label{eq:poten}
\zq
where $\rho^2=\sigma^2+\tau^2+\pi^2$,
and now we introduced explicitly the ultraviolet cutoff $\Lambda$.

By using the integral representation for the generalized Riemann zeta
function $\zeta$ \cite{22},
\xq
\int^\infty_0 ds s^{\mu-1} e^{-\beta s} \coth s  = \Gamma(\mu)
\left[2^{1-\mu}\zeta (\mu,\frac{\beta}{2})-\beta^{-\mu}\right],\label{eq:zeta}
\zq
which is valid at $\mu>1$, and analytically continuing this representation
to $\mu=-\frac{1}{2}$, we can rewrite Eq.~(\ref{eq:poten}) as
\beq
V(\rho) &=& \frac{N}{\pi} \Bigg[\frac{\Lambda}{2\sqrt{\pi}}
\left(\frac{\sqrt{\pi}}{g}-1\right) \rho^2-\frac{\sqrt{2}}{l^3}\zeta
\left(-\frac{1}{2}, \frac{(\rho l)^2}{2}+1\right) - \nonumber\\
&-& \frac{\rho}{2l^2}\Bigg]+O(1/\Lambda),     \label{eq:poten41}
\eeq
where  the magnetic length $l$ is $l=|eB|^{-1/2}$ and here we introduced
the dimensionless coupling constant
\xq
g\equiv N\frac{\Lambda}{\pi}G.       \label{eq:gG}
\zq
(We recall that $\zeta(\mu,\frac{\beta}{2})$ is defined as
\xq
\zeta (\mu, \frac{\beta}{2})=\sum^\infty_{n=0}\frac{1}{(n+\frac{\beta}{2})^\mu}
\zq
at $\mu>1$ \cite{22}.)

Let us now analyze the gap equation $dV/d\rho=0$. It is
\xq
\frac{\Lambda\rho}{\pi g}=\frac{\rho}{2\pi^{3/2}} \int^\infty_{1/\Lambda^2}
\frac{ds}{s^{3/2}} e^{-s\rho^2}eBs\coth(eBs), \label{eq:gap1}
\zq
which can be rewritten as
\beq
2\Lambda l(\frac{1}{g}-\frac{1}{\sqrt{\pi}})\rho &=& \frac{1}{l} +\sqrt{2} \rho
\zeta(\frac{1}{2},1+\frac{\rho^2l^2}{2})+O(1/\Lambda).\label{eq:gap2}
\eeq
As $B\to0$, we recover the known gap equation \cite{14}:
\xq
\rho^2=\rho\Lambda\left(\frac{1}{\sqrt{\pi}}-\frac{1}{g}\right).\label{eq:gap}
\zq
It admits a nontrivial solution only if the coupling constant $g$ is
supercritical, $g>g_c=\sqrt{\pi}$ (as Eq.~(\ref{eq:lag2}) implies, a solution
to the gap
equation, $\sigma=\bar{\sigma}$, coincides with the fermion dynamical mass,
$\bar{\sigma}=m_{dyn}$). We will show that the magnetic field changes the
situation dramatically: at $B\neq0$, a nontrivial solution exists at all $g>0$.
The reason for this is that the magnetic field enhances the interaction in the
infrared region (large $s$): at $B\neq0$, the integral in Eq.~(\ref{eq:gap1})
becomes proportional to $1/\rho$ as $\rho\to0$.

We shall first consider the case of subcritical $g$, $g<g_c=\sqrt{\pi}$, which
in turn can be divided into two subcases: a) $g\ll g_c$ and b) $g\to g_c-0$
(nearcritical $g$). Assuming that $|\bar{\sigma}l|\ll1$ at $g\ll g_c$, we find
from Eq.~(\ref{eq:gap2}):
\xq
m_{dyn}\equiv\bar{\sigma}\simeq\frac{|eB|g\sqrt{\pi}}{2\Lambda(g_c-g)}.
\label{eq:mdyn}
\zq
Since Eq.~(\ref{eq:mdyn}) implies that the condition $|\bar{\sigma}l|\ll1$
fulfills at all
$g$ satisfying $(g_c-g)\gg\frac{|eB|^{1/2}}{\Lambda}$, the relation
(\ref{eq:mdyn}) is actually valid in that whole region.

Note the following interesting point. Eq.(\ref{eq:cons}) implies that
$m_{dyn}=\langle 0|\sigma|0\rangle=-\pi g/N\Lambda
\langle 0|\bar{\Psi}\Psi|0\rangle$. From here and Eq.(\ref{eq:mdyn}) we find
that the condensate $\langle 0|\bar{\Psi}\Psi|0\rangle$ is
$\langle 0|\bar{\Psi}\Psi|0\rangle=-N|eB|/2\pi$ in leading order in $g$,
{\em i.e.} it coincides with the value of the condensate calculated in the
problem of free fermions in a magnetic field (see Eq.(\ref{eq:condensate})).
This point implies that at small $g$ (weakly interacting fermions) the
$| \theta_1,\theta_2\rangle$--vacua
constructed in Sec.3 are good trial states for the vacua of the problem with
interacting fermions. This point also explains why the dynamical mass
$m_{dyn}$ in this problem is an analytic function of $g$ at $g=0$: indeed,
the condensate exists already at $g=0$!

At $g_c-g\ltwid\sqrt{|eB|}/\Lambda$, introducing the scale
$m^*=\Lambda(1/g-1/g_c)$, we get the equation
\xq
2m^*l=\frac{1}{|\sigma|l}+\sqrt{2}\zeta \left(\frac{1}{2}, \frac{(\sigma
l)^2}{2}
+1\right)
\zq
which implies that in the nearcritical region, $m_{dyn}$ is
\xq
m_{dyn}=\bar{\sigma}\sim|eB|^{1/2}. \label{eq:mdynB}
\zq
Thus in the scaling region, with $g_c-g\ltwid\sqrt{|eB|}/\Lambda$, the
cutoff disappears from the observable $m_{dyn}$. This agrees with the
well-known fact that the critical value $g_c=\sqrt{\pi}$ is an ultraviolet
stable fixed point at leading order in $1/N$ \cite{14}.
 The relation (\ref{eq:mdynB}) can
be considered as a scaling law in the scaling region.

In the supercritical region, at $g>g_c$, the analytic expression for $m_{dyn}$
can be obtained at weak $|eB|$, satisfying the condition
$\sqrt{|eB|}/m^{(0)}_{dyn}\ll 1$, where $m^{(0)}_{dyn}$ is the solution
of the gap equation (\ref{eq:gap}) with $B=0$. Then, using the asymptotic
formula~\cite{22}
\xq
\zeta (z,q) \overrightarrow{q\to\infty} \frac{1}{(z-1)q^{z-1}}
\left[1+\frac{z-1}{2q}+\dots\right], \label{eq:asym}
\zq
we find from Eq.~(\ref{eq:gap2})
\xq
m_{dyn}=\bar{\sigma}=m_{dyn}^{(0)}\left[1+\frac{(eB)^2}{12(m^{(0)}_{dyn})^4}
\right] ,  \label{eq:md51}
\zq
i.e., $m_{dyn}$ increases with $B$. The numerical study of the equation
(\ref{eq:gap2})
shows that $m_{dyn}$ increases with $B$ at all values of $g$ and $B$.

A striking fact is that, unlike the gap equation (\ref{eq:gap}) with $B=0$,
the gap
equation with $B\neq0$ does not have the trivial solution $\sigma=0$. Indeed,
Eq.~(\ref{eq:poten}) implies that
$dV/d\sigma|_{\sigma=0}=d\tilde{V}/d\sigma|_{\sigma=0}$, and
then we find from Eqs.~(\ref{eq:lag2}) and (\ref{eq:exp}) that
\xq
\frac{d\tilde{V}}{d\sigma}|_{\sigma=0}=\langle0|\bar{\Psi}\Psi|0\rangle |_{g=0}
=-N\frac{|eB|}{2\pi}\neq 0
\zq
(see Eq.~(\ref{eq:condensate})). Thus, despite the spontaneous
character of the $U(2)$ symmetry breakdown, there
is no trivial solution (stable or unstable) in the magnetic field at all values
of $g$.

\section {The NJL Model in a Magnetic Field. The Kinetic Term in the Effective
Action.}

Let us now consider the kinetic term ${\cal L}_k$ in the effective action
(\ref{eq:effact}).

The $U(2)$ symmetry implies that the general form of ${\cal L}_k$ is
\beq
{\cal L}_k &=& N\frac{F_1^{\mu\nu}}{2} (\partial_\mu\rho_j\partial_\nu \rho_j)
+ N\frac{F^{\mu\nu}_2}{\rho^2}(\rho_j\partial_\mu\rho_j)\cdot\nonumber\\
\qquad &\cdot& (\rho_i\partial_\nu\rho_i) \label{eq:Lk}
\eeq
where $\mbox{\boldmath$\rho$}
=(\sigma,\tau,\pi)$ and $F^{\mu\nu}_1$, $F^{\mu\nu}_2$ are
functions of $\rho^2=\sigma^2+\tau^2+\pi^2$. To find the functions
$F^{\mu\nu}_1$, $F^{\mu\nu}_2$, one can use different methods. We used the
method of Ref.~\cite{23}. The derivation of ${\cal L}_k$ is considered in
Appendix~A. Here we shall present the final results.

The functions $F^{\mu\nu}_1$ and $F^{\mu\nu}_2$ take the form
$F^{\mu\nu}_1=g^{\mu\nu}F^{\mu\mu}_1$, $F^{\mu\nu}_2=g^{\mu\nu}F^{\mu\mu}_2$
where
\beq
F^{00}_1 &=& \frac{l}{8\pi} \left(\frac{1}{\sqrt{2}}\zeta \Big(\frac{3}{2},
  \frac{(\rho l)^2}{2}+1\Big)+(\rho l)^{-3}\right), \nonumber\\
F^{11}_1 &=& F^{22}_1=\frac{1}{4\pi\rho}, \nonumber\\
F^{00}_2 &=&- \frac{l}{16\pi} \left(\frac{(\rho l)^2}{2\sqrt{2}} \zeta
  \Big(\frac{5}{2}, \frac{(\rho l)^2}{2}+1\Big) + (\rho l)^{-3}\right),
  \label{eq:Fij} \\
F_2^{11} &=& F^{22}_2 = \frac{l}{8\pi} \Bigg[\frac{(\rho l)^4}{\sqrt{2}} \zeta
   \Big(\frac{3}{2}, \frac{(\rho l)^2}{2}+1\Big) + \nonumber\\
\qquad &+& \sqrt{2}(\rho l)^2 \zeta \Big(\frac{1}{2},
   \frac{(\rho l)^2}{2}+1\Big) +2\rho l-(\rho l)^{-1}\Bigg]\nonumber
\eeq
(we recall that the magnetic length $l\equiv|eB|^{-1/2}$).

We would like to emphasize that, as follows from Eq.(\ref{eq:Fij}), the
propagator of the NG modes in leading order in $1/N$ has a genuine
$(2+1)$--dimensional form. We shall see in Appendix~C that this fact is
crucial for providing the reliability of the $1/N$ expansion in this
problem (physical reasons for the $(2+1)$--dimensional character of the
dynamics of the neutral NG bosons are considered in the next section).

Now, knowing the effective potential and the kinetic term, we can define the
energy
spectrum (dispersion law) of the collective excitations $\sigma$ and
$\tau,\pi$.

\section {The NJL Model in a Magnetic Field. The Spectrum of the Collective
Excitations.}

We begin by considering the spectrum of the collective excitations in the
subcritical, $g<g_c$, region.

At $g_c-g\gg\sqrt{|eB|}/\Lambda$ (where $|\bar{\sigma}l|\ll1$, see
Eq.~(\ref{eq:mdyn})), we find from Eqs.~(\ref{eq:Lk}) and (\ref{eq:Fij})
the dispersion law for the $\tau$
and $\pi$ NG (gapless) modes:
\xq
E_{\tau,\pi}\simeq\sqrt{2}(\bar{\sigma}l)(
\mbox{\boldmath$k$}^2)^{1/2} =
\frac{gg_c|eB|^{1/2}}{\sqrt{2}\Lambda(g_c-g)}
(\mbox{\boldmath$k$}^2)^{1/2}    \label{eq:72}
\zq
(see Eq.~(\ref{eq:mdyn})). As the interaction is switched off, $g\to0$,
their velocity, $v=gg_c(g_c-g)^{-1}\sqrt{|eB|/2\Lambda^2}$,
becomes zero, and we
return to the dynamics with spontaneous flavor symmetry breaking but without
genuine NG modes discussed in Section~3.

In order to define the ``mass" (energy gap) $M_\sigma$ of the $\sigma$-mode, we
note that
\beq
\frac{d^2V}{d\sigma^2}|_{\sigma=\bar{\sigma}} &=& N
\frac{\bar{\sigma}^2l}{\pi^{3/2}} \int^\infty_0ds \sqrt{s}
\exp (-(\bar{\sigma}l)^2s)
\coth s=\nonumber\\
&=& N\frac{\bar{\sigma}^2l}{2\pi} \left[\frac{1}{\sqrt{2}}\zeta
\Big(\frac{3}{2},
\frac{(\bar{\sigma}l)^2}{2}+1\Big)+(\bar{\sigma}l)^{-3}\right] \label{eq:d2V}
\eeq
(see Eq.~(\ref{eq:poten})). Then we find from Eqs.~(\ref{eq:Lk}),
(\ref{eq:Fij}) and (\ref{eq:d2V}) that
\xq
M^2_\sigma \simeq \frac{8\sqrt{2}(g_c-g)}{gg_c\zeta(\frac{3}{2})}
\Lambda |eB|^{1/2}
\zq
at $g_c-g\gg\sqrt{|eB|}/\Lambda$. As $g\to0$, the $\sigma$--mode
decouples $(M_\sigma\to\infty)$.

Thus the dynamics in the problem of a relativistic fermion in
an external magnetic field emerges from this model in the limit when the
interaction between fermions is switched off. The attractive $(g>0)$
interaction ``resurrects" the NG modes and they acquire a velocity $v\sim g$.

Let us now consider the nearcritical region with
$g_c-g\ltwid\sqrt{|eB|}/\Lambda$. From Eqs.~(\ref{eq:Lk}),
(\ref{eq:Fij}), we find that
\xq
E_{\tau,\pi}=f(\bar{\sigma}l)(\mbox{\boldmath$k$}^2)^{1/2},\label{eq:75}
\zq
where
\beq
f(\bar{\sigma}l)=(\frac{2}{\bar{\sigma}l})^{1/2} \left(\frac{1}{\sqrt{2}} \zeta
\Big(\frac{3}{2}, \frac{(\bar{\sigma}l)^2}{2}+1\Big)+ (\bar{\sigma}l)^{-3}
\right)^{-1/2}.\label{eq:ffun}
\eeq
Since in this nearcritical (scaling) region the parameter $\bar{\sigma}$ is
$\bar{\sigma}\sim|eB|^{1/2}=l^{-1}$, we conclude that the cutoff $\Lambda$
disappears from the observables $E_\tau$ and $E_\pi$ in the scaling region.

In the same way, we find from Eqs.~(\ref{eq:Lk}),
(\ref{eq:Fij}) and (\ref{eq:d2V}) that
\xq
M^2_\sigma\sim |eB|
\zq
in the scaling region.

Let us turn to the supercritical region with $g>g_c$. The analytic expressions
for $E_{\tau,\pi}$ and $M^2_\sigma$ can be obtained for small $|eB|=l^{-2}$,
satisfying the condition $|\bar{\sigma}l|\gg1$. Then, using the asymptotic
formula (\ref{eq:asym}) for zeta functions, we find from Eqs.~(\ref{eq:Lk}),
(\ref{eq:Fij}) and (\ref{eq:d2V}):
\beq
E_{\tau,\pi} &=& \left(1-\frac{1}{8(\bar{\sigma}l)^4}\right)
(\mbox{\boldmath$k$}^2)^{1/2},\label{eq:78} \\
M_\sigma^2 &=& 6\bar{\sigma}^2\left(1-\frac{3}{4}
\frac{1}{(\bar{\sigma}l)^2}\right),
\eeq
where $\bar{\sigma}$ is given in Eq.~(\ref{eq:md51}). These relations show that
the
magnetic field leads to decreasing both the velocity of the NG modes (it
becomes less than 1) and the mass (energy gap) of the $\sigma$ mode.

Let us indicate the following interesting point intimately connected with
the $(2+1)$--dimensional character of the dynamics of the neutral NG
modes. The $(2+1)$--dimensional character is reflecded in that the velocity
${\bf v}_{\pi,\tau}=\partial E_{\pi,\tau}/\partial {\bf k}$ is not zero.
As follows from Eqs.(\ref{eq:72}), (\ref{eq:75}) and (\ref{eq:78}), the
velocity ${\bf v}_{\pi,\tau}$ decreases with $m_{dyn}=\bar{\sigma}$ and
becomes zero ({\em i.e.} the dynamics becomes $(0+1)$--dimensional) when
$m_{dyn}\to 0$, {\em i.e.} when the interaction is switched off ($g\to 0$).
The reason for this is clear: since at $g=0$ the energy of the neutral
system made up  of a fermion and an antifermion from the lowest Landau
level is identically zero, its velocity is also zero. This fact in turn
reflects the point that the motion of charged fermions in the $x_1$-$x_2$
plane is restricted by a magnetic field. On the other hand, at $g>0$, there
are genuine neutral NG bound states (with the bound energy $\Delta
E_{\pi,\tau}\equiv 2m_{dyn}-E_{\pi,\tau}|_{k=0}=2m_{dyn}$). Since the motion
of the center of the mass of {\em neutral} bound states is not restricted
by a magnetic field, their dynamics is $(2+1)$--dimensional.

Let us now discuss the continuum limit $\Lambda\to\infty$ in
more detail. As is known, at $B=0$, in this model, an interacting continuum
theory appears only at the critical value $g=g_c =\sqrt\pi$ (the continuum
theory is trivial at $g<g_c$) \cite{4,14,Kondo,15}. Therefore, since
at $g<g_c$, in the continuum limit, there is no attractive interaction
between fermions, it is not surprising that at $g<g_c$, the dynamical mass
$m_{dyn}\sim g|eB|/\Lambda$ disappears as $\Lambda\to\infty$.

At $B=0$, the continuum theory is in the symmetric phase at $g\to g_{c}-0$
and in the broken phase at $g\to g_{c}+0$ . On the other hand, as follows
from our analysis, in a magnetic field , it is in the broken phase both at
$g\to g_{c}-0$ and $g\to g_{c}+0$ (though the dispersion relations for
fermions and collective excitations $\rho$ are different at $g\to g_{c}-0$
and $g\to g_{c}+0$ ).

Up to now we have considered four-component fermions. In the case of
two-component fermions, the effective potential, $V_2$ is
$V_2(\sigma)=V(\sigma)/2$ where $V(\sigma)$ is defined in
Eqs.~(\ref{eq:poten}) and
(\ref{eq:poten41}). However, the essential new point is that there is no
continuous $(U(2))$
symmetry (and therefore NG modes) in this case. As in the case of
four-component fermions, in an external magnetic field, the dynamical fermion
mass (now breaking parity) is generated at any positive value of the coupling
constant $g$.

The NJL model illustrates the general phenomenon in $2+1$ dimensions: in the
infrared region, a magnetic field reduces the dynamics of fermion pairing to a
one-dimensional dynamics (at the lowest Landau level), thus catalysing the
generation of a dynamical mass for fermions. A concrete sample of dynamical
symmetry breaking is of course different in different models.

\section {Thermodynamic Properties of the NJL Model in a Magnetic Field.}

In this section, we will study the thermodynamic properties of the NJL model in
a magnetic field. In particular, we will show that there is a symmetry
restoring phase transition at high temperature.

Our goal is to determine the thermodynamic (effective) potential in the NJL
model in a magnetic field. Although we are mostly interested in studying the
system at finite temperature $T$ and zero chemical potential  $\mu$ (i.e., at
equal densities of fermions and antifermions), we shall derive the effective
potential $V_{\beta,\mu}(\sigma)$ (at the leading order in $1/N$) at
arbitrary values of $\beta=1/T$ and $\mu$.

Since in the leading order in $1/N$, the effective potential $V(\sigma)\equiv
V_{\beta,\mu}(\sigma)|_{\stackrel{{\textstyle\beta=\infty}}
{{\textstyle \!\!\mu=0}}}$ is
given by a sum of one-(fermion)loop diagrams, the thermodynamic potential is in
this approximation:
\beq
V_{\beta,\mu}(\sigma) &=& V(\sigma)+\tilde{V}_{\beta,\mu}(\sigma)=\nonumber\\
&=& \frac{N}{\pi} \left[\frac{\Lambda}{2\sqrt{\pi}} (\frac{\sqrt{\pi}}{g}-1)
\sigma^2-\frac{\sqrt{2}}{l^3}\zeta \Big(-\frac{1}{2}, \frac{(\sigma l)^2}{2}
+1\Big) -\frac{\sigma}{2l^2}\right]-\nonumber\\
&-& N \frac{|eB|}{2\pi\beta} \Bigg\{\ln (1+e^{-\beta(\sigma-\mu)})
+2\sum_{k=1}^\infty\ln\Big(1+ \nonumber\\
&+& e^{-\beta(\sqrt{\sigma^2+\frac{2k}{l^2}}-\mu)\Big)}
+(\mu\to-\mu)\Bigg\}\label{eq:temperature}
\eeq
(see Eq.~(\ref{eq:poten41})). Here the sum is taken over all the fermion and
antifermion
(with $\mu\to-\mu$) Landau levels; the factor $|eB|/2\pi$ describes the
degeneracy of each level. The relation (\ref{eq:temperature}) is
derived (in the framework of the
imaginary time formalism \cite{24}) in Appendix~B. We also show there that it
can be rewritten as
\beq
V_{\beta,\mu}(\sigma) &=& \frac{\sigma^2}{2G} +\frac{N}{4\pi^{3/2}l^3}
\int^\infty_0 \frac{dt}{t^{3/2}} e^{-(tl^2\sigma^2)} \coth t\cdot \nonumber\\
&\cdot& \Theta_4\left(\frac{i}{2}\mu\beta|\frac{i}{4\pi t}
(\frac{\beta}{l})^2\right),\label{eq:jacob}
\eeq
where $\Theta_4$ is the fourth Jacobian theta function \cite{22}.

Henceforth we will consider the case of zero chemical potential corresponding
to equal densities of fermions and antifermions in the system. In this case,
the thermodynamic potential $V_\beta \equiv V_{\beta,\mu}|_{\mu=0}$ is:
\beq
V_\beta(\sigma) &=& \frac{N}{\pi} \Bigg[\frac{\Lambda}{2\sqrt{\pi}}
(\frac{\sqrt{\pi}}{g}-1) \sigma^2- \frac{\sqrt{2}}{l^3}\zeta \Big(-\frac{1}{2},
\frac{(\sigma l)^2}{2}+1\Big) - \nonumber\\
&-& \frac{\sigma}{2l^2}\Bigg]  - N \frac{|eB|}{\pi\beta}
\Bigg[\ln(1+e^{-\beta\sigma})+ \nonumber\\
\qquad &+& 2 \sum^\infty_{k=1} \ln
\Big(1+e^{-\beta\sqrt{\sigma^2+\frac{2k}{l^2}}}\Big)\Bigg].
\eeq

We solved numerically the gap equation
\xq
\frac{dV_\beta(\sigma)}{d\sigma}=0.
\zq
The main result is that at $T=T_c\sim m_{dyn}$ (actually at $T=T_c\simeq
m_{dyn/2}$), there is a symmetry restoring (second order) phase transition
(see Figs.~1 and 2). The phase diagram in the $B-T$ plane is shown in Figs.
{}~3 and 4.

We recall that there cannot be spontaneous breakdown of a continuous symmetry
at finite $(T>0)$ temperature in $2+1$ dimensions (the Mermin-Wagner-Coleman
(MWC) theorem \cite{20}). This happens because at nonzero temperature
the dynamics of a zero mode in $2+1$ dimensional field theories is two
dimensional. As a result, strong fluctuations of would be NG modes lead to
vanishing the order parameter connected with a spontaneous breakdown of a
continuous symmetry. In the NJL model with a finite temperature
(both at $B=0$ and in a magnetic field),
the MWC theorem manifests itself only beyond the leading order in $1/N$.
One plausible possibility of what happens at $T\neq 0$ beyond the leading
order in $1/N$ is the
following.
The dynamics of the zero mode in this model is essentially equivalent
to that of the $SU(2)$ $\sigma$--model in $2$ dimensional Euclidean
space. As is known, the $SU(2)$ symmetry is exact in the latter model
and, as a result, the would be NG bosons become massive exitations
\cite{Polyakov}. Therefore it seems plausible that in the
$(2+1)$--dimensional NJL model in a magnetic field, the $SU(2)$ symmetry
will be restored at any finite temperature, and the dynamically generated
mass $m_{dyn}$ of fermions will disappear.

The question whether this, or another, scenario is realized at finite
temperature in this model deserves further study.

\section {Conclusion}

The main result of this paper is that a magnetic field is a strong catalyst,
generating a fermion mass (energy gap), in $2+1$ dimensions. It would be worth
considering the present effect in $2+1$ dimensional effective  theories
describing high temperature superconductivity and the quantum Hall effect
where a magnetic field is an important ingredient of the dynamics.In
connection with this, we note that in some models of high temperature
superconductivity of Ref.~\cite{2}, the energy gap in the electron spectrum
results from electron-hole (i.e., fermion--antifermion rather than
fermion--fermion) pairing. Also, using the four--component spinors in
these models reflects the presence of two sublattices in high temperature
superconductors.
Another, potentially interesting, application of the present effect may be in
$3+1$ dimensional field theories at high temperature. Since at high
temperature, their dynamics effectively reduces to that of $2+1$ dimensional
theories, it might happen that in a magnetic field, at high temperature,
fermions (quarks in quark--gluon plasma, for example) acquire a dynamical mass
and NG excitations appear.

It would be interesting to check the realization of this effect in $2+1$
dimensional theories in lattice computer simulations. Note that the recent
computer simulations of the $2+1$
dimensional NJL model \cite{15} show that the $1/N$ expansion is quite
reliable, at least at $N\geq12$.

The essence of the present effect is that in a constant magnetic field, the
dynamics of fermion pairing is one-dimensional: the pairing takes place
essentially for fermions at the (degenerate) lowest Landau level. This implies
the universal character of this effect in $2+1$ dimensions.

In this paper, we considered the dynamics in the presence of a constant
magnetic field only. It would be interesting to extend this analysis to the
case of inhomogeneous electromagnetic fields in $2+1$ dimensions.
In connection with this we note that the present effect is intimately connected
with the fact that in $2+1$ dimensions, the massless Dirac equation in a
constant magnetic field admits an infinite number of normalized solutions with
$E=0$ (zero modes); more precisely, the density of such solutions is finite.
One may expect that the same effect will take place for any electromagnetic
field configuration in which the density of zero modes is finite.
As we have known recently, the program of the derivation of a low energy
effective
action in $2+1$ dimensional QED in external electromagnetic fields has been
developed in Ref.~\cite{26}.

V.A.M. is grateful to the members of the Department of Applied Mathematics
of the University of Western Ontario,
where part of this work was done, and of the Institute for Theoretical Physics
of the University of California (Santa Barbara)  for their hospitality. He
thanks J.M.~Cornwall, E.~D'~Hoker, D.~Kaplan, J.~Polchinski, S.~Raby,
L.~Randall, J.~Schwarz, A.~Vainshtein, L.C.R.~Wijewardhana, K.~Yamawaki, and
A.~Zee for useful discussions.

The research was supported in part by the National Science Foundation under
Grant No. PHY89-04035 and by the International Soros Science Education
Program (ISSEP) through grant No.PSU052143.

\medskip

{\Large{\bf Appendix A}}

In this Appendix, we derive the expressions for the fermion propagator
and for the kinetic term ${\cal L}_k$ in the effective action.

In the coordinate space, the fermion propagator
is
\beq
S(x,y) &=& (i\hat{D}+m)_x\langle
x|\frac{-i}{m^2+\hat{D}^2}|y\rangle=\nonumber\\
&=& (i\hat{D}+m)_x\int^\infty_0ds\langle x|\exp[-is(m^2+\hat{D}^2)]|y\rangle,
\label{eq:Sxy}
\eeq
where $\hat{D}\equiv\gamma^\mu D_\mu$ and $D_\mu$ is the covariant derivative
in Eq.~(\ref{eq:dmu}).

The matrix element $\langle x|e^{-is(m^2+\hat{D}^2)}|y\rangle$ can be
calculated by using the Schwinger (proper time) approach \cite{16}. It is
\beq
\langle x|e^{-is(m^2+\hat{D}^2)}|y\rangle &=&
  \frac{e^{-i\frac{\pi}{4}}}{8(\pi s)^{3/2}} e^{i[S_{cl}-sm^2]}
\Big(eBs\cot(eBs)+\label{eq:xy}\\
\qquad&+& \gamma^1\gamma^2eBs\Big),\nonumber
\eeq
where
\beq
S_{cl} &=& e \int^x_y A_\lambda^{ext} dz^\lambda - \frac{1}{4s} (x-y)_\nu
\Big[g^{\nu\mu}+\frac{((F^{ext})^2)^{\mu\nu}}{B^2} \Big(1-\nonumber\\
\qquad&-& eBs\cot(eBs)\Big)\Big](x-y)_\mu.
\eeq
Here the integral is calculated along the straight line.

{}From Eqs.~(\ref{eq:Sxy}) and (\ref{eq:xy}), we find the expression
(\ref{eq:green}) for the fermion propagator.

Let us now consider the derivation of the kinetic term (\ref{eq:Lk})
in the low-energy effective action:
\beq
{\cal L}_k &=& N \frac{F_1^{\mu\nu}}{2} (\partial_\mu\rho_j\partial_\nu\rho_j)
+N \frac{F^{\mu\nu}_2}{\rho^2} (\rho_j\partial_\mu\rho_j)\cdot \nonumber\\
\qquad &\cdot& (\rho_i\partial_\nu\rho_i),\label{eq:Lkapend}
\eeq
where $\bfrho=(\sigma,\tau,\pi)$ and $F^{\mu\nu}_1$, $F^{\mu\nu}_2$ depend
on the $U(2)$-invariant $\rho^2=\sigma^2+\tau^2+\pi^2$. The definition
$\Gamma_k=\int d^3x{\cal L}_k$ and Eq.~(\ref{eq:Lkapend}) imply that the form
of the
functions $F^{\mu\nu}_1$, $F^{\mu\nu}_2$ is determined from the equations:
\beq
 N^{-1}\frac{\delta^2\Gamma_k}{\delta\sigma(x)\delta\sigma(0)}
 |_{\stackrel{{\textstyle\sigma=const}}{{\textstyle\tau=\pi=0}}} &=&-
  (F^{\mu\nu}_1+2F^{\mu\nu}_2)
|_{\stackrel{{\textstyle\sigma=const}}{{\textstyle\tau=\pi=0}}}
  \cdot \partial_\mu\partial_\nu\delta^3(x), \\
 N^{-1}\frac{\delta^2\Gamma_k}{\delta\pi(x)\delta\pi(0)}
 |_{\stackrel{{\textstyle\sigma=const}}{{\textstyle\tau=\pi=0}}}
    &=&- F^{\mu\nu}_1
 |_{\stackrel{{\textstyle\sigma=const}}{{\textstyle\tau=\pi=0}}}
  \cdot\partial_\mu\partial_\nu\delta^3(x).\label{eq:d2G}
\eeq
Here $\Gamma_k$ is the part of the effective action (\ref{eq:effact})
containing terms with two derivatives. Eq.~(\ref{eq:effact}) implies that
$\Gamma_k=\tilde{\Gamma}_k$.  Therefore we find from Eq.~(\ref{eq:d2G}) that
\beq
F^{\mu\nu}_1 &=&-\frac{N^{-1}}{2} \int d^3xx^\mu x^\nu
\frac{\delta^2\tilde{\Gamma}_k}{\delta\pi(x)\delta\pi(0)}
=-\frac{N^{-1}}{2} \int
d^3 x x^\mu x^\nu \cdot\nonumber\\
&\cdot& \frac{\delta^2\tilde{\Gamma}}{\delta\pi(x)\delta\pi(0)}
\eeq
(henceforth we shall not write explicitly the condition $\tau=\pi=0$,
$\sigma=const$). Taking into account the definition of the fermion
propagator,
\xq
iS^{-1}=i\hat{D}-\sigma,
\zq
we find from Eq.~(\ref{eq:tildefact}) that
\beq
\frac{\delta^2\tilde{\Gamma}}{\delta\pi(x)\delta\pi(0)} &=&- itr
\Big(S(x,0)i\gamma^5S(0,x)i\gamma^5\Big)=\nonumber\\
&=&-itr\Big(\tilde{S}(x,0)i\gamma^5\tilde{S}(0,x)i\gamma^5\Big) =\nonumber\\
&=&-i\int\frac{d^3kd^3q}{(2\pi)^6}e^{iqx} tr
\Big(\tilde{S}(k)i\gamma^5\tilde{S}(k+q)i\gamma^5\Big)
\eeq
(the functions $\tilde{S}(x)$ and $\tilde{S}(k)$ are given in
Eqs.~(\ref{eq:green})--(\ref{eq:20})).
Therefore,
\xq
F^{\mu\nu}_1=-\frac{iN^{-1}}{2}\int \frac{d^3k}{(2\pi)^3}tr
\Bigg(\tilde{S}(k)i\gamma^5\frac{\partial^2\tilde{S}(k)}{\partial k_\mu\partial
k_\nu}i\gamma^5\Bigg).             \label{eq:F1}
\zq

In the same way, we find that
\beq
F^{\mu\nu}_2 &=&-\frac{iN^{-1}}{4} \int \frac{d^3k}{(2\pi)^3}tr
\Big(\tilde{S}(k)\frac{\partial^2\tilde{S}(k)}{\partial k_\mu\partial
k_\nu}\Big)-\frac{1}{2}F^{\mu\nu}_1=\label{eq:F2}\\
&=&-\frac{iN^{-1}}{4}\int\frac{d^3k}{(2\pi)^3}tr
\Bigg(\tilde{S}(k)\frac{\partial^2\tilde{S}(k)}{\partial k_\mu\partial k_\nu} -
\tilde{S}(k)i\gamma^5 \frac{\partial^2\tilde{S}(k)}{\partial k_\mu\partial
k_\nu}i\gamma^5\Bigg).\nonumber
\eeq

Taking into account the expression for $\tilde{S}(k)$ in Eq.~(\ref{eq:20})
(with $m=\sigma$), we get:
\beq
\frac{\partial^2\tilde{S}(k)}{\partial k^0\partial k^0} &=& 2il^4 \int^\infty_0
dtt\exp ((R(t))\{\sigma(1+\eta\gamma^1\gamma^2T)+\nonumber\\
&+& 3k^0\gamma^0(1+\eta\gamma^1\gamma^2T)-k^i\gamma^i(1+T^2)+ 2itl^2(k^0)^2
\sigma(1+\\
&+& \eta\gamma^1\gamma^2T)+2itl^2(k^0)^3\gamma^0(1+\eta\gamma^1\gamma^2T)-
 2itl^2(k^0)^2(k^i\gamma^i)(1+T^2)\},\nonumber\\
\frac{\partial^2\tilde{S}(k)}{\partial k^j\partial k^j}&=&-2il^4 \int^\infty_0
dt T\exp(R(t))\{\sigma(1+\eta\gamma^1\gamma^2T)-\nonumber\\
&-& k^i\gamma^i(1+T^2)-2k^j\gamma^j(1+T^2)
+k^0\gamma^0(1+\eta\gamma^1\gamma^2T) -\nonumber\\
&-&2iTl^2(k^j)^2\sigma (1+\eta\gamma^1\gamma^2T)-2iTl^2(k^j)^2 k^0\gamma^0(1+
\nonumber\\
&+& \eta \gamma^1\gamma^2T)+2iTl^2(k^j)^2k^i\gamma^i(1+T^2)\}\label{eq:d2S}
\eeq
($i,j=1,2$; there is no summation over $j$), where
\beq
\eta&=& sign (eB), \quad T=\tan t,\nonumber\\
R(t) &=&-it(\sigma l)^2+it(k^0)^2-il^2 \mbox{\boldmath$k$}^2 T
\eeq

Eqs.~(\ref{eq:20}), (\ref{eq:F1}), (\ref{eq:F2}) imply that nondiagonal
terms $F^{\mu\nu}_1$ and $F^{\mu\nu}_2$ are equal to zero.
The diagonal terms are determined from
Eqs.~(\ref{eq:20}), (\ref{eq:F1})--(\ref{eq:d2S}),
after rather long, although routine, calculations:
\beq
F^{00}_1 &=& \frac{l}{12\pi^{3/2}} \int^\infty_0 d\tau
\frac{\sqrt{\tau}}{\sinh\tau} e^{-(\sigma l)^2\tau} \Big[(\sigma l)^2\tau
\cosh\tau +\frac{3}{2}\cosh\tau+\nonumber\\
&+& \frac{\tau}{\sinh\tau}\Big] =\frac{l}{8\pi}
\Bigg(\frac{1}{\sqrt{2}}\zeta\Big(\frac{3}{2}, \frac{(\sigma l)^2}{2} +1\Big) +
(\sigma l)^{-3}\Bigg),\\
F^{00}_2 &=&- \frac{l(\sigma l)^2}{12\pi^{3/2}} \int^\infty_0 d\tau
\tau^{3/2}e^{-(\sigma l)^2\tau} \coth\tau = \nonumber\\
&=&- \frac{l}{16\pi} \Bigg(\frac{(\sigma l)^2}{2\sqrt{2}} \zeta
\Big(\frac{5}{2}, \frac{(\sigma l)^2}{2}+1\Big) +(\sigma
l)^{-3}\Bigg),\\
F^{11}_1 &=& F^{22}_1=\frac{1}{4\pi\sigma},\\
F^{11}_2 &=& F^{22}_2 =\frac{l(\sigma l)^2}{4\pi^{3/2}} \int^\infty_0 d\tau
\tau^{-1/2} e^{-(\sigma l)^2\tau} \coth \tau (1-\tau\coth\tau) = \nonumber\\
&=& \frac{l}{8\pi} \Bigg(\frac{(\sigma l)^4}{\sqrt{2}}\zeta \Big(\frac{3}{2},
\frac{(\sigma l)^2}{2}+1\Big)+ \sqrt{2}(\sigma l)^2 \zeta \Big(\frac{1}{2},
\frac{(\sigma l)^2}{2}+1\Big) +\nonumber\\
&+& 2\sigma l-(\sigma l)^{-1}\Bigg).
\eeq
Here (besides Eq.~(\ref{eq:zeta})) the following relations were used \cite{22}:
\beq
\int^\infty_0 \frac{\tau^{\mu-1}e^{-\beta\tau}}{\sinh^2\tau}d\tau &=& 2^{1-\mu}
\Gamma(\mu)\Big[2\zeta(\mu-1,\frac{\beta}{2})-\nonumber\\
\qquad &-&\beta\zeta (\mu, \frac{\beta}{2})\Big], \mu>2,\\
\int^\infty_0 \tau^{\mu-1}e^{-\beta\tau}\coth^2\tau d\tau &=&
\beta^{-\mu}\Gamma(\mu) + \int^\infty_0
\frac{\tau^{\mu-1}e^{-\beta\tau}}{\sinh^2\tau} d\tau,\mu>2, \\
\int^\infty_0\frac{\tau^{\mu-1}e^{-\beta\tau}\coth\tau}{\sinh^2\tau}d\tau &=&
\frac{\mu-1}{2} \int^\infty_0
 \frac{\tau^{\mu-2}e^{-\beta\tau}}{\sinh^2\tau}d\tau
-\nonumber\\
&-&\frac{\beta}{2}\int^\infty_0
\frac{\tau^{\mu-1}e^{-\beta\tau}}{\sinh^2\tau}d\tau, \mu>3.
\eeq

{\Large{\bf Appendix B}}

In this Appendix we shall derive the thermodynamic potential $V_{\beta,\mu}$
in the NJL model (\ref{eq:lag1}); here $\beta=1/T$ is an inverse temperature
and $\mu$ is a chemical potential.

As is well known \cite{24}, the partition function
\xq
Z_{\beta,\mu}=Tr\Big[\exp(-\beta H')\Big]
\zq
is expressed through a path integral over fields of a system (here
$H'=H-\mu\int\bar{\Psi}\gamma^0\Psi d^2x$, $H$ is the Hamiltonian of the
system). In the NJL model (\ref{eq:lag1}), (\ref{eq:lag2}),
the path integral is:
\beq
Z_{\beta,\mu} &=& \int [d\Psi] [d\bar{\Psi}] [d\sigma][d\tau][d\pi] \exp
\Big\{ i\int^{-i\beta}_0dt\int d^2x \cdot\nonumber\\
&\cdot& \Big[\bar{\Psi}iS^{-1}\Psi-\frac{1}{2G} \rho^2\Big]\Big\}
\eeq
where $\rho^2=\sigma^2+\tau^2+\pi^2$, $S$ is the fermion propagator
(\ref{eq:green}) with $m$ replaced by $\sigma+\gamma^3\tau+i\gamma^5\pi$,
and while the fermion fields satisfy the antiperiodic boundary conditions,
\xq
\Psi|_{t=0}=-\Psi|_{t=-i\beta},\quad
\bar{\Psi}|_{t=0}=-\bar{\Psi}|_{t=-i\beta},\label{eq:bound}
\zq
the boson fields satisfy the periodic boundary conditions.

In order to calculate the thermodynamic potential $V_{\beta,\mu}(\rho)$, it
is sufficient to consider configurations with $\tau=\pi=0$ and $\sigma=const$.
Then the potential is defined as:
\beq
\exp \Big\{-\beta V_{\beta,\mu}[\int d^2x]\Big\}
&=& \int[d\Psi][d\bar{\Psi}]\exp \Big\{i\int^{-i\beta}_0dt\int
d^2x\cdot\nonumber\\
&\cdot& \Big[\bar{\Psi}iS^{-1}\Psi-\frac{1}{2G}\sigma^2\Big]\Big\}.
\eeq
At the leading order in $1/N$, this potential defines the thermodynamic
properties of the system.

As is known \cite{24}, in the formalism of the imaginary time, the
thermodynamic potential $V_{\beta,\mu}$ can be obtained from the
representation for the effective potential $V$, at $T=0$ and $\mu=0$,
by replacing
\beq
\int\frac{d^3k}{(2\pi)^3} &\to & \frac{i}{\beta}\sum^{+\infty}_{n=-\infty}
\int\frac{d^2k}{(2\pi)^2}, \nonumber\\
k^0 &\to & i\omega_n+\mu;\quad \omega_n=\frac{\pi}{\beta}(2n+1)
\eeq
($\omega_n=\frac{\pi}{\beta}(2n+1)$ follows from the antiperiodic conditions
(\ref{eq:bound})). Then, using the representation for the effective potential
in Section~4
and the expression (\ref{eq:20}) for the fermion propagator, we get:
\beq
V_{\beta,\mu}(\sigma) &=& \frac{\sigma^2}{2G}+\frac{N}{2\pi\beta l^2}
\int^\infty_0\frac{dt}{t} e^{-tl^2(\sigma^2-\mu^2)}\coth t \cdot\nonumber\\
&\cdot& \Theta_2\Big(2\pi t\frac{\mu l^2}{\beta}| 4i\pi t\frac{l^2}{\beta^2}
\big),   \label{eq:pot}
\eeq
where
\xq
\Theta_2(u|\tau)=2\sum^\infty_{n=0} e^{i\pi\tau(n+\frac{1}{2})^2} \cos
\Big((2n+1)u\Big)
\zq
is the second Jacobian theta function \cite{22}.

By using the identity \cite{22}
\xq
\Theta_2(u|\tau)=(\frac{i}{\tau})^{1/2} e^{-i\frac{u^2}{\pi\tau}} \Theta_4
\Big(\frac{u}{\tau}| -\frac{1}{\tau}\Big),
\zq
where
\xq
\Theta_4(u|\tau)=1+2\sum^\infty_{n=1}(-1)^n e^{i\pi n^2\tau}\cos(2nu)
\zq
is the fourth Jacobian theta function, one can rewrite the relation
(\ref{eq:pot}) as
\beq
V_{\beta,\mu}(\sigma) &=& \frac{\sigma^2}{2G} +\frac{N}{4\pi^{3/2}l^3}
\int^\infty_0\frac{dt}{t^{3/2}}  e^{-tl^2\sigma^2}\coth t \cdot\nonumber\\
&\cdot& \Theta_4\Big(\frac{i}{2}\mu\beta|\frac{i}{4\pi
t}(\frac{\beta}{l})^2\Big)=\label{eq:Vbmu}\\
&=& V(\sigma)+\frac{N}{2\pi^{3/2}l^3}\int^\infty_0 \frac{dt}{t^{3/2}} \coth
t\sum^\infty_{n=1}(-1)^n\cosh (\mu\beta\ n)\cdot\nonumber\\
\qquad &\cdot& \exp-(t\sigma^2l^2+\frac{\beta^2n^2}{4tl^2}),\nonumber
\eeq
where $V(\sigma)$ is the effective potential (\ref{eq:poten}).

Thus we have derived the representation (\ref{eq:jacob}) for the
thermodynamic potential. Let us show that it is equivalent to
the representation (\ref{eq:temperature}).

By using the series
\xq
\coth t=1+2\sum^\infty_{m=1} e^{-2tm},
\zq
the expression for $\tilde{V}_{\mu,\beta}=V_{\mu,\beta}-V$ in
Eq.~(\ref{eq:Vbmu}) can be rewritten as
\beq
\tilde{V}_{\mu,\beta} &=& \frac{N}{\pi l^2\beta}\sum^\infty_{n=1} (-1)^n
\frac{\cosh(\mu\beta n)}{n}  \Big[e^{-\beta\sigma n}+\\
&+& 2\sum^\infty_{m=1}\exp (-\beta\sigma n \sqrt{1+\frac{2m}{(\sigma l)^2}})
\Big].\nonumber
\eeq
Here we also used the relations \cite{22}:
\beq
\int^\infty_0dx x^{\nu-1}\exp (-\frac{\beta}{x}-\gamma x) &=& 2
(\frac{\beta}{\gamma})^{\nu/2} K_\nu (2\sqrt{\beta\gamma}), \nonumber\\
K_{-\frac{1}{2}} (z) = K_{\frac{1}{2}} (z) &=& (\frac{\pi}{2z})^{1/2} e^{-z},
\eeq
where $K_\nu(z)$ is a modified Bessel function.

Since
\xq
\sum^\infty_{n=1} (-1)^n\frac{e^{\alpha n}+e^{-\alpha n}}{n} e^{-\beta n} =-
\ln(1+e^{-2\beta}+2e^{-\beta}\cosh\alpha),
\zq
we find that
\beq
\tilde{V}_{\mu,\beta} &=&- \frac{N}{2\pi\beta l^2} \Big\{\ln
[1+e^{-2\beta\sigma}+2e^{-\beta\sigma} \cosh (\mu\beta)\Big] + \\
&+& 2 \sum^\infty_{m=1} \ln
\Big[1+e^{-2\beta\sigma\sqrt{1+\frac{2m}{(\sigma l)^2}}} +2
 e^{-\beta\sigma\sqrt{1+\frac{2m}{(\sigma l)^2}}} \cosh(\mu\beta)\Big]\Big\}.
\nonumber
\eeq
It is now easy to check that the expression for the thermodynamic potential
$V_{\beta,\mu}=V+\tilde{V}_{\beta,\mu}$ coincides with that in
Eq.~(\ref{eq:temperature}).

{\Large{\bf Appendix C}}

In this Appendix we analyze the next--to--leading order in $1/N$ expansion
in the $(2+1)$--dimensional NJL model at zero temperature. Our main goal
is to show that the propagator of the neutral NG bosons $\pi$ and $\tau$
have a $(2+1)$--dimensional structure in this approximation and that (unlike
the $(1+1)$--dimensional Gross--Neveu model \cite{19}) the $1/N$
expansion is reliable in this model.

A review of the $1/N$ expansion in $(2+1)$--dimensional four--fermion
interaction models can be found in Ref.\cite{14}. For our
purposes, it is sufficient to know that this perturbative expansion is
given by Feynman diagrams with the vertices and the propagators of
fermions and composite particles $\sigma$, $\pi$ and $\tau$ calculated
in leading order in $1/N$. In leading order, the fermion propagator
is given in Eqs.(\ref{eq:green})--(\ref{eq:21}).
As follows from Eq.(\ref{eq:lag2}), the Yukawa coupling of
fermions with $\sigma$, $\tau$ and $\pi$ is $g_Y=1$ in this
approximation. The inverse propagators of $\sigma$, $\tau$ and $\pi$
are \cite{14,23}:
\beq
D^{-1}_{\mbox{\boldmath $\rho$}}(x)=N\left(\frac{\Lambda}{g\pi}\delta^3(x)
+itr[S(x,0)T_{\mbox{\boldmath $\rho$}}S(0,x)T_{\mbox{\boldmath $\rho$}}]
\right),
\eeq
where $\mbox{\boldmath $\rho$}=(\sigma,\tau,\pi)$ and $T_{\sigma}=1$,
$T_{\tau}=\gamma^3$,
$T_{\pi}=i\gamma^5$. Here $S(x,0)$ is the fermion propagator
(\ref{eq:green}) with the mass $m_{dyn}=\bar{\sigma}$ defined
from the gap equation (\ref{eq:gap2}). For completeness, we
write down the explicit expression for the Fourier transform of
the propagators of the NG bosons:
\beq
D^{-1}_{\tau}(k)&=&D^{-1}_{\pi}(k)=\frac{N}{4\pi^{3/2}l}\int\limits_{0}^{1}du
\int\limits_{0}^{\infty}\frac{ds\sqrt{s}}{\sinh{s}}\exp[-s(\sigma l)^2]
\nonumber\\
&\cdot&\Bigg[
\bigg(1-\exp[R(s,u)]\bigg)\left((\sigma l)^2\cosh{s}+
\frac{1}{\sinh{s}}-\frac{\cosh{s}}{s} \right)\nonumber\\
&&-2(lk_0)^2(1-u^2)\cosh{s}\exp[R(s,u)]+3l^2{\bf k}^2\exp[R(s,u)]\nonumber\\
&\cdot&\left(\cosh{su}-u\sinh{su}\coth{s}+
\frac{2}{3\sinh^2{s}}(\cosh{su}-\cosh{s})\right)\Bigg],
\eeq
where
\beq
R(s,u)&\stackrel{def}{=}&\frac{s}{4}(lk_0)^2(1-u^2)-
\frac{l^2{\bf k}^2}{2}\frac{\cosh{s}-\cosh{su}}{\sinh{s}}.
\eeq
Actually, for our purposes, we need to know the form of these
propagators at small momenta only. We find from
Eqs.(\ref{eq:Lk}), (\ref{eq:Fij}):
\beq
D_{\tau}(k)=D_{\pi}(k)=-\frac{4\pi\bar{\sigma}}{N}f^2(\bar{\sigma}l)
[k_0^2-f^2(\bar{\sigma}l){\bf k}^2]^{-1}
\label{eq:NG}
\eeq
where
\beq
f(\bar{\sigma}l)=\left(\frac{2}{\bar{\sigma}l}\right)^{1/2}
\left(\frac{1}{\sqrt{2}}\zeta\left(\frac{3}{2},\frac{(\bar{\sigma}l)^2}{2}
+1\right)+(\bar{\sigma}l)^{-3}\right)^{-1/2}
\eeq
(see Eq.(\ref{eq:ffun})).

The crucial point for us is that, because of the dynamical mass $m_{dyn}$,
the fermion propagator is soft in the infrared region
(see Eq.(\ref{eq:poles})) and that the propagators of the $\tau$ and $\pi$
(\ref{eq:NG}) have a $(2+1)$--dimensional form in the infrared region (as
follows from Eqs.(\ref{eq:Lk}), (\ref{eq:Fij}) the propagator of $\sigma$
has of course also a $(2+1)$--dimensional form).

Let us begin by considering the next--to--leading corrections in the
effective potential. The diagram which contributes to the effective
potential in this order is shown in Fig.5a. Because of the structure
of the propagators pointed out above, there are no infrared divegences
in this contribution to the potential. (Note that this is in contrast
to the Gross--Nevue model: because of a $(1+1)$--dimensional form of
the propagators of the NG bosons, this contribution is logarithmically
divergent in the infrared region in that model, {\em i.e.} the
$1/N$ expansion is unreliable in that case).
Therefore the diagram in Fig.5a leads to a finite, $O(1)$, correction to
the potential $V$ (we recall that the leading contribution in $V$ is of
order N). As a result, at sufficiently large values of N, the gap
equation in next--to--leading order in $1/N$ in this model admits a
nontrivial solution $\bar{\rho}\neq 0$. Since the potential depends only on
the radial variable $\rho$, the angular variables $\theta$ and $\varphi$
($\mbox{\boldmath $\rho$}=(\rho\cos\theta,
\rho\sin\theta\cos\varphi, \rho\sin\theta\sin\varphi)$),
connected with the $\tau$ and $\pi$, appear
in the effective lagrangian only through their derivatives. This in turn
implies that the $\tau$ and $\pi$ retain to be gapless NG modes in the
next--to--leading order in $1/N$.

Let us now consider the next--to--leading corrections to the propagators of
these NG modes. First of all, note that in a constant magnetic field, the
propagator of a neutral local field $\varphi(x)$, $D_{\varphi}(x,y)$, is
translation invariant, {\em i.e.} it depends on $(x-y)$. This immediately
follows from the fact that the operators of space translations (\ref{eq:tran})
take the canonical form for neutral fields (the operator of time
translations is $i\partial/\partial t$ for both neutral and charged
fields in a constant magnetic field).  The diagrams contributing to the
propagators of the NG modes in this order are shown in Fig.5b. Because
of the dynamical mass $m_{dyn}$ in the fermion propagator, this contribution
is analytic at $k_{\mu}=0$. Since at large $N$ the gap equation has a
nontrivial solution in this approximation, there is no contribution of
$O(k^0)\sim const$ in the inverse propagators of $\tau$ and $\pi$.
Therefore the first term in the monentum expansion of this contribution
has the form $C_1k_0^2-C_2{\bf k}^2$, where $C_1$ and $C_2$ are functions
of $\bar{\sigma}l$, {\em i.e.} the propagators take the following form
in this approximation:
\beq
D_{\tau}(k)=D_{\pi}(k)\stackrel{k\to 0}{=}
-\frac{4\pi\bar{\sigma}}{N}f^2(\bar{\sigma}l)\left[\left(1-\frac{1}{N}
\tilde{C}_1(\bar{\sigma}l)\right)k_0^2-\left(f^2(\bar{\sigma}l)-
\frac{1}{N}\tilde{C}_2(\bar{\sigma}l)\right){\bf k}^2\right]^{-1}.
\eeq
(see Eq.(\ref{eq:NG})).

Because of the same reasons, there are also no infrared divergences either
in the fermion propagator (see Fig.5c) or in the Yukawa vertices
(see Fig.5d) in this order. Therefore at sufficiently large values of $N$,
the results retain essentially the same as in leading order in $1/N$.

We believe that there should not be principal obstacles to extend this
analysis for all orders in $1/N$.

\newpage

{\Large{\bf Figure Captions}}

Fig. 1. The thermodynamic potential $V_\beta$ as a function of $\sigma$ at
different temperatures at subcritical $g$: $g/g_c=0.9$. All quantities
are measured in $\mu\equiv\Lambda/g_c$ units: $V_\beta\to\pi V_\beta/\mu^3$,
$eB\to eB/\mu^2$, $\sigma\to\sigma/\mu$.

Fig. 2. The thermodynamic potential $V_\beta$ as a function of $\sigma$ at
different temperatures at supercritical $g$: $g/g_c=10^3$. All quantities
are measured in $\mu\equiv\Lambda/g_c$ units: $V_\beta\to\pi V_\beta/\mu^3$,
$eB\to eB/\mu^2$, $\sigma\to\sigma/\mu$.

Fig. 3. The critical line in the $eB-T$ plane separating the asymmetric $(A)$
and  symmetric
$(S)$ phases at subcritical $g$: $g/g_c=0.9$. All quantities are measured
in $\mu\equiv\Lambda/g_c$ units: $eB\to eB/\mu^2$, $T\to T/\mu$.

Fig. 4. The critical line in the $eB-T$ plane separating the asymmetric $(A)$
and symmetric $(S)$ phases at supercritical $g$: $g/g_c=1.1$. All quantities
are measured in $\mu=\Lambda/g_c$ units: $eB\to eB/\mu^2$, $T\to T/\mu$.

Fig. 5. Diagrams in next--to--leading order in $1/N$. A solid line denotes
the fermion propagator and a dashed line denotes the propagators of $\sigma$,
$\tau$ and $\pi$ in leading order in $1/N$.


\begin{thebibliography}{99}

\bibitem{1} R. Jackiw, {\sl Phys. Rev.} {\bf D29} (1984) 2375; I. Affleck, {\sl
Nucl. Phys.} {\bf B265} (1986) 409; I.V. Krive and A.S. Rozhavskii, {\sl Usp.
Fiz. Nauk} {\bf 152} (1987) 33 [{\sl Sov. Phys. Usp.} {\bf 30} (1987) 370].

\bibitem{2} A. Kovner and B. Rosenstein, {\sl Phys. Rev.} {\bf B42} (1990)
4748; G.W. Semenoff and L.C.R. Wijewardhana, {\sl Phys. Rev.} {\bf D45}
 (1992) 1342;  N. Dorey and N.E. Mavromatos,
 {\sl Nucl. Phys.} {\bf B368} (1992) 614; R.
MacKenzie, P.K. Panigrahi and S. Sakhi, {\sl Phys. Rev.} {\bf B48} (1993) 3892.

\bibitem{3} J. Bardeen, L.N. Cooper and J.R. Schrieffer {\sl Phys. Rev.} {\bf
108} (1957) 1175.

\bibitem{4} V.A. Miransky, {\sl Dynamical Symmetry Breaking in Quantum Field
Theories\/} (World Scientific, Singapore, 1993).

\bibitem{5} Y. Nambu and G. Jona-Lasinio, {\sl Phys. Rev.} {\bf 122} (1961)
345.


\bibitem{6} I.V. Krive and S.A. Naftulin, {\sl Phys. Rev} {\bf D46} (1992)
2737.

\bibitem{7} A.I. Akheizer and V.B. Berestetsky, {\em Quantum Electrodynamics}
(Interscience, New York, 1965).

\bibitem{8} T. Appelquist, M. Bowick, D. Karabali, and L.C.R. Wijewardhana,
 {\sl  Phys. Rev.} {\bf D33} (1986) 3704.

\bibitem{9} T. Appelquist, D. Nash and L.C.R. Wijewardhana, {\sl Phys. Rev.
Lett.\/} {\bf 60} (1988) 2575.

\bibitem{10} E. Dagotto, J.B. Kogut, and A. Koci\'{c}, {\sl Phys. Rev. Lett.\/}
{\bf 62} (1989) 1083; {\sl Nucl. Phys.} {\bf B334} (1990) 279; S. Hands, and
J.B. Kogut, {\sl Nucl. Phys.} {\bf B335} (1990) 455.

\bibitem{11} D. Atkinson, P.W. Johnson, and P. Maris, {\sl Phys. Rev.} {\bf
D42} (1990) 602; K.-I. Kondo and H. Nakatani, {\sl Mod. Phys. Lett.} {\bf A5}
(1990) 407; {\sl Prog. Theor. Phys.} {\bf 87} (1992) 193.

\bibitem{12} M.R. Pennington and D. Walsh, {\sl Phys. Lett.} {\bf B253} (1991)
246;  R. Pisarski, {\sl Phys. Rev.} {\bf D44} (1991) 1866.

\bibitem{13} M.C. Diamantini, G.W. Semenoff, and P. Sodano, {\sl Phys. Rev.
Lett.} {\bf 70} (1993) 3438.

\bibitem{14} B. Rosenstein, B.J. Warr and S.H Park, {\sl Phys. Rep.} {\bf 205}
(1991) 59.

\bibitem{Kondo}K.I. Kondo, M. Tanabashi and K. Yamawaki,
{\sl Prog. Theor. Phys.} {\bf 89} (1993) 1249.

\bibitem{15} S. Hands, A. Koci\'{c}, and J.B. Kogut, {\sl Ann. Phys} {\bf 224}
(1993) 29.

\bibitem{16} J. Schwinger, {\sl Phys. Rev.} {\bf 82} (1951) 664.

\bibitem{22} I.S. Gradshtein and I.M. Ryzhik, {\em Table of Integrals, Series
and Products} (Academic Press, Orlando, 1980).

\bibitem{Cho} A. Chodos, K. Everding, and D.A. Owen, {\sl Phys. Rev.} {\bf D42}
(1990) 2881.

\bibitem{17} P. Lee and S. Shenoy, {\sl Phys. Rev. Lett.} {\bf 28} (1972) 1025.
See also Z. Tesanovic, {\sl Physica} {\bf C220} (1994) 303 for an update and
the list of references.

\bibitem{Haag} R. Haag, {\sl Nuovo Cimento} {\bf 25} (1962) 287.

\bibitem{Niemi} See the first paper in Ref.1; A.J. Niemi and G.W. Semenoff,
{\sl Phys. Rev. Lett.} {\bf 51} (1983) 2077; {\sl Phys. Rep.} {\bf 135}
(1986) 99; D. Boyanovsky, R. Blankenbecler and R. Yahalom, {\sl Nucl. Phys.}
{\bf B270} (1986) 483.

\bibitem{Kovner} A. Kovner, B. Rosenstein and D. Eliezer, {\sl Nucl. Phys.}
{\bf B350} (1991) 235; A.Kovner and D. Eliezer, {\sl Int. J. Mod. Phys.}
{\bf A7} (1992) 2775.

\bibitem{18} N.N. Bogolyubov, {\sl Physica} {\bf 26} (1960) 1; R.V. Lange,
{\sl Phys. Rev.} {\bf 146} (1966) 301.

\bibitem{19} D.J. Gross and A.Neveu, {\sl Phys. Rev.} {\bf D10} (1974) 3225;
E. Witten, {\sl Nucl. Phys.} {\bf B145} (1978) 110.

\bibitem{20} N.D. Mermin and H. Wagner, {\sl Phys. Rev. Lett.} {\bf 17} (1966)
1133; S. Coleman, {\sl Commun. Math. Phys.} {\bf 31} (1973) 259.

\bibitem{Zak} J. Zak, {\sl Phys. Rev.} {\bf 134} (1964) A1602; J.E. Avron,
I.W. Herbst and B. Simon, {\sl Ann. Phys.} {\bf 114} (1978) 431.

\bibitem{23} V.P. Gusynin and V.A. Miransky, {\sl Mod. Phys. Lett.}
{\bf A6} (1991) 2443; {\sl Sov. Phys. JETP} {\bf 74} (1992) 216;
V.A. Miransky, {\sl Int. J. Mod. Phys.} {\bf A8} (1993) 135.

\bibitem{24} L. Dolan and R. Jackiw, {\sl Phys. Rev.} {\bf D9} (1974) 3320.

\bibitem{Polyakov} A.M.Polyakov and P.B.Wiegmann, {\sl Phys. Lett.} {\bf 131B}
(1983) 121.

%\bibitem{25} V.L. Berezinsky, {\sl Sov. Phys. JETP} {\bf 32} (1970) 493;
%J.M. Kosterlitz and D.J. Thouless, {\sl J. Phys.} {\bf C6} (1973) 1181.

\bibitem{26} D. Cangemi, E. D'Hoker and G.Dunne,{\sl Phys. Rev.} {\bf D51}
(1995) R2513.

\end{thebibliography}
\end{document}